\PassOptionsToPackage{hyphens}{url}
\PassOptionsToPackage{obeyspaces}{url}
\PassOptionsToPackage{spaces}{url}
\PassOptionsToPackage{dvipsnames}{xcolor}
\documentclass[conference,compsoc]{IEEEtran}

\usepackage[dvipsnames]{xcolor}

\usepackage[english]{babel}
\usepackage[nolist,nohyperlinks]{acronym}
\usepackage{multirow}
\usepackage{wasysym}
\usepackage{subcaption}
\usepackage{graphicx}
\usepackage{url}
\usepackage{hyperref}

\usepackage[final]{changes}
\usepackage{tikz}
\newcommand*\circled[1]{\tikz[baseline=(char.base)]{
            \node[shape=circle,draw,inner sep=1pt,semithick] (char) {\small #1};}}

\newcommand{\del}[1]{}

\newcommand{\ntg}[1]{}

\usepackage{xparse}
\usepackage{xstring}
\NewDocumentCommand{\rfc}{m o}{
    \StrCount{#1}{,}[\commacount]
    \ifnum\commacount > 0
        \def\rfcprefix{RFC}
    \else
        \def\rfcprefix{RFC}
    \fi
    \IfValueTF{#2}{
        \StrCount{#2}{,}[\commacount]
        \ifnum\commacount < 1
            \def\secprefix{\S}
        \else
            \def\secprefix{\S\S}
        \fi
        \rfcprefix#1\secprefix#2
    }{
        \rfcprefix#1
    }
}

\usepackage{array}
\makeatletter
\newcolumntype{"}{@{\hskip\tabcolsep\vrule width 1pt\hskip\tabcolsep}}
\newcolumntype{;}{@{\hskip\tabcolsep\vrule width 1pt}}
\makeatother
\usepackage{tikz}
\usepackage{textcomp}
\usepackage{hyperref}
\usepackage{lipsum}

\newcommand\copyrighttext{
  \footnotesize \textcopyright 2026 IEEE. Personal use of this material is permitted. Permission from IEEE must be obtained for all other uses, in any current or future media, including reprinting/republishing this material for advertising or promotional purposes, creating new collective works, for resale or redistribution to servers or lists, or reuse of any copyrighted component of this work in other works.
}
\newcommand\copyrightnotice{
\begin{tikzpicture}[remember picture,overlay]
\node[anchor=south,yshift=10pt] at (current page.south) {\fbox{\parbox{\dimexpr\textwidth-\fboxsep-\fboxrule\relax}{\copyrighttext}}};
\end{tikzpicture}
}

\begin{document}

\begin{acronym}
    \acro{rpki}[RPKI]{Resource Public Key Infrastructure}
    \acroindefinite{rpki}{an}{a}

    \acro{bgp}[BGP]{Border Gateway Protocol}
    
    \acro{rrdp}[RRDP]{\acs{rpki} Repository Delta Protocol}
    \acroindefinite{rrdp}{an}{a}
    
    \acro{rtr}[RTR]{\acs{rpki} to Router Protocol}
    \acroindefinite{rtr}{an}{a}
    
    \acro{rpsl}[RPSL]{Routing Policy Specification Language}
    \acroindefinite{rpsl}{an}{a}
    
    \acro{rp}[RP]{Relying Party}
    \acroplural{rp}[RPs]{Relying Parties}
    \acroindefinite{rp}{an}{a}

    \acro{rfc}[RFC]{Request for Comments}
    \acroplural{rfc}[RFCs]{Requests for Comments}

    \acro{roa}[ROA]{Route Origin Authorization}
    \acroindefinite{roa}{an}{a}
    
    \acro{cms}[CMS]{Cryptographic Message Syntax}
    
    \acro{crl}[CRL]{Certificate Revocation List}
    
    \acro{ca}[CA]{Certificate Authority}
    \acroplural{ca}[CAs]{Certificate Authorities}
    
    \acro{as}[AS]{Autonomous System}
    \acroindefinite{as}{an}{an}
    
    \acro{pp}[PP]{Publication Point}
    
    \acro{rir}[RIR]{Regional Internet Registry}
    \acroplural{rir}[RIRs]{Regional Internet Registries}
    \acroindefinite{rir}{an}{a}
    
    \acro{lir}[LIR]{Local Internet Registry}
    \acroplural{lir}[LIRs]{Local Internet Registries}
    
    \acro{slurm}[SLURM]{Simplified Local Internet Number Resource Management}
    
    \acro{tal}[TAL]{Trust Anchor Locator}
    
    \acro{cp}[CP]{Certificate Policy}
    \acroplural{cp}[CPs]{Certificate Policies}
    
    \acro{rov}[ROV]{Route Origin Validation}
    \acroindefinite{rov}{an}{a}
    
    \acro{vrp}[VRP]{Validated \acs{roa} Payload}
    
    \acro{dos}[DoS]{denial-of-service}

    \acro{gbr}[GBR]{Ghostbuster Record}
\end{acronym}

\title{The Fault in Our Drafts:\\Vulnerabilities in RPKI Specification and Software}

\author{\IEEEauthorblockN{Oliver Jacobsen\IEEEauthorrefmark{1}\IEEEauthorrefmark{2},
Tobias Kirsch\IEEEauthorrefmark{1}\IEEEauthorrefmark{2},
Haya Schulmann\IEEEauthorrefmark{1}\IEEEauthorrefmark{2}, 
Niklas Vogel\IEEEauthorrefmark{1}\IEEEauthorrefmark{2} and
Michael Waidner\IEEEauthorrefmark{1}\IEEEauthorrefmark{3}\IEEEauthorrefmark{4}}
\IEEEauthorblockA{\IEEEauthorrefmark{1}ATHENE -- National Research Center for Applied Cybersecurity, Germany}
\IEEEauthorblockA{\IEEEauthorrefmark{2}Goethe-Universität Frankfurt, Germany}
\IEEEauthorblockA{\IEEEauthorrefmark{3}Fraunhofer SIT, Germany}
\IEEEauthorblockA{\IEEEauthorrefmark{4}TU Darmstadt, Germany}}

\IEEEoverridecommandlockouts
\makeatletter\def\@IEEEpubidpullup{0\baselineskip}\makeatother

\maketitle
\copyrightnotice

\begin{abstract}
The \ac{rpki} secures the Internet’s routing system by defining a complex trust and validation framework for certificates, Route Origin Authorizations (ROAs), manifests, and Certificate Revocation Lists (CRLs).
These mechanisms are specified across dozens of RFCs.

This paper presents the first comprehensive analysis of the causal link between flaws in \ac{rpki} \acp{rfc} and vulnerabilities in implementations and real-world deployments.
We reveal how vague, conflicting, or underspecified requirements in 50 RPKI RFCs propagate into inconsistent implementation behavior and operational failures.

We conduct the first large-scale, impact-driven evaluation of \ac{rpki} specifications. 
Our methodology combines differential fuzzing of major RPKI implementations with Internet-wide crawling and validation log analysis, enabling us to trace practical vulnerabilities back to flawed RFC requirements.
We uncover 61 previously undocumented inconsistencies in validation behavior, trace 23 directly to RFC flaws, and identify two novel vulnerabilities that were assigned CVEs. 

Our findings reveal that these are not isolated coding errors but rather systemic issues inherent in how RPKI standards are written, interpreted, and implemented. 
To mitigate these threats, we propose concrete recommendations and introduce a novel alerting service that monitors and reports live inconsistencies in RPKI deployments. 
Our open-source datasets, code, and tools support reproducibility and further research.
\end{abstract}

\IEEEpeerreviewmaketitle

\section{Introduction}
The \ac{bgp} is insecure and vulnerable to prefix hijacks, exposing systems to attacks like espionage, theft of cryptocurrency, outages, distribution of malware~\cite{bgp:attacks,DBLP:journals/cacm/SunABVRCM21,klayswap, cloudfarehijack}.
\ac{rpki} was designed to protect \ac{bgp} from prefix hijacks
by certifying assigned network resources and allowing routers to validate BGP announcements against the certified resources~\cite{DBLP:journals/cacm/SunABVRCM21}.
Already, more than 55\% of the Internet prefixes are certified with RPKI~\cite{nistmonitor}, and about 30\% of networks validate BGP announcements with RPKI~\cite{rovistameasurement}.
With the growing adoption of RPKI, it also becomes central to Internet resilience and stability, and is critical for national security~\cite{whitehouse2024routing}. 
The recent routing security roadmap of the White-House identifies RPKI as a mature, ready-to-implement technology to mitigate vulnerabilities in BGP and recommends its deployment across all networks \cite{whitehouse2024routing}.
Nevertheless, no comprehensive analysis of RPKI's resilience and security has been performed to establish its maturity and readiness for real-world adoption; the complexity of RPKI makes such analysis difficult.

{\bf \ac{rpki} is complex.} 
To ease implementation and improve security (\rfc{6480}), the design of \ac{rpki} builds on existing concepts and technologies, such as X.509 certificates, incremental update mechanisms, and a hierarchical tree structure, implementing the global distribution of routing security information.

This design is specified in over 50 RFC documents, making \ac{rpki} one of the most complex cryptography-based infrastructures deployed at Internet scale, with multiple object types, validation workflows, and integration with third-party systems.
This complexity makes it hard to trace requirements across documents and identify specification-level issues that may lead to divergent or insecure implementations.
In the past, research has found vulnerabilities in RPKI, but these have been viewed in isolation, attributed to mere implementation bugs, deviations from the standard, lack of developer tools, or connectivity issues~\cite{gilad2017we,hlavacek2020disco,van2022rpkiller, hlavacek2022stalloris,mirdita2023cure,mirdita2024sok}, not to systemic flaws in specification. 

{\bf Study of systemic RFC issues.}
In this work, we perform the first study of the ca. 50 RPKI RFCs and show that bugs, inconsistencies, and vulnerabilities in {\em standard-compliant} RPKI implementations are caused by flaws {\em within RFCs}.
We classify these flaws according to their underlying root causes, deriving systematic problems in the RFCs.

In addition to the 50 \ac{rpki} RFCs analysis, our study encompasses an in-depth examination of all popular RPKI validator software packages and an investigation of the operational practices and deployments across the Internet.
We trace 24 out of the 52 previously known RPKI vulnerabilities directly to flaws in the standard RFC documents, demonstrating that these are not isolated incidents resulting from poor implementation or operational practices, nor are they the result of deviations from the standard requirements.  
We find that 7 out of 32 {\em registered CVEs} in RPKI were caused by flawed requirements in the RFCs. 
We identify 12 issues in the specification, all of which manifest in real-world implementations, and 2 vulnerabilities that expose RPKI deployments to attacks.
We also find 61 {\em inconsistencies} in RPKI implementations, 23 of which can be directly attributed to the aforementioned issues in RFCs.
Our analysis reveals that past investigations of RPKI inconsistencies have not been correctly contextualized within the RFCs; as a result, the underlying specification issues have not been addressed.
One example is an issue in the extensions within resource certificates that caused 6K Amazon prefixes not to be validated.
While the developers attempted different solutions in their implementations, the core issue is still present in the RFC, as we explain in \S\ref{sc:termdef}.

{\bf Methodology to identify practical issues.} 
Our methodology is designed to uncover specification flaws in RPKI RFCs by observing how they manifest in real-world implementations and deployments, which is a key strength of our approach. 
Rather than analyzing RFCs in isolation or deriving tests solely from their logic, we begin by identifying practical inconsistencies, vulnerabilities, and deviations observed in the wild. 
We then trace these issues back to ambiguities, contradictions, or omissions in the relevant standards. 
To identify such issues, we combine a systematic analysis of RPKI software with an investigation of Internet-wide deployments. 
We use two techniques: (1) a customized differential fuzzing framework to expose discrepancies in validator behavior caused by undefined or conflicting edge cases in the RFCs, and (2) a novel crawler that detects operational inconsistencies and anomalies in \ac{rpki} repositories, which often stem from under-specified requirements. 
The findings from both approaches are combined with issues from public sources, such as CVE reports, GitHub, and related scientific publications. 
This provides a comprehensive view of both novel and previously reported RPKI problems, which we leverage for identifying their root causes within the specifications. 

Harnessing practical impact as the primary method for identifying flaws offers advantages over solely relying on specification text, as in formal methods-based approaches~\cite{pacheco2022automated, okumura2020formal, yen2021sage}. 
Certain types of issues, such as undefined edge cases, vague language, or under-specified interactions between components, are often difficult to detect by analyzing RFC text in isolation. 
In contrast, our methodology identifies issues that manifest in real-world software or deployments, ensuring that each finding reflects tangible operational risks rather than a purely theoretical concerns that lack practical impact, such as formatting errors or typos.

\textbf{Real-world impact.}
Evaluating the real-world impact of identified issues presents its own challenges.
The RPKI ecosystem comprises a large and evolving dataset, currently exceeding 400K objects, deployed in around $\sim$100 repositories with large variance in size, configurations, and response behavior.
Currently, no tools exist for systematically analyzing these deployments.
We develop a crawler that downloads and compares real-world objects, checks them against RFC specification and the inconsistencies we identify in our analysis, and evaluates their real-world impact.

\indent {\bf Mitigations and alert tool.} 
We provide recommendations for resolving the issues identified in this work and for improving the RFCs. 
We notified members of the IETF working group for Secure Interdomain Routing Operations (SIDROPS)\footnote{\url{https://datatracker.ietf.org/wg/sidrops/about/}} about these issues and are in active discussions on how to resolve them in the specification. Additionally, we develop a web service with an alert tool for the community that notifies users of current problems in RPKI in the wild (\url{https://rpki-notify.site}). 
The tool constantly monitors the live RPKI infrastructure, provides information on newly occurring issues, and alerts operators in cases of unreachable publication points, erroneous RPKI objects, and inconsistencies between the \acp{rp}. 
Furthermore, we use the insights gained through our empirical analysis to explore how LLM-based tools could assist future research in identifying flawed requirements in RFCs. 
Based on the types of issues we encountered in our practical analysis, the LLM tool receives a keyword or requirement as input and retrieves the relevant sections from the RFCs.
This enables a broader application of our findings for performing future semi-automated specification reviews; for example, during the creation of new RFCs.
We open-source our compiled datasets, test cases, and implementations to aid developers in mitigating the issues.\footnote{\url{https://github.com/tkh42/rfc-analysis}}

\indent {\bf Organization.}  
We review RPKI in \S\ref{sec:background} and Related work in \S\ref{sc:works}. 
Our methodology is in \S\ref{sc:methodology}. 
In \S\ref{sc:rfc-analysis}, we identify RFC flaws and their propagation into RPKI implementations, and in \S\ref{sc:inconsistencies} their impact on RPKI validation. 
In \S\ref{sc:discussion} we attribute the issues to core design principles of RPKI and explain their real-world impact. 
Mitigations, conclusions, and ethics are in \S\ref{sc:mitigations}, \S\ref{sc:conclusion}, and \S\ref{sc:ethics} respectively.
 
\section{Overview of RPKI}\label{sec:background}

\begin{figure}[t!]
    \centering
    \includegraphics[width=\linewidth]{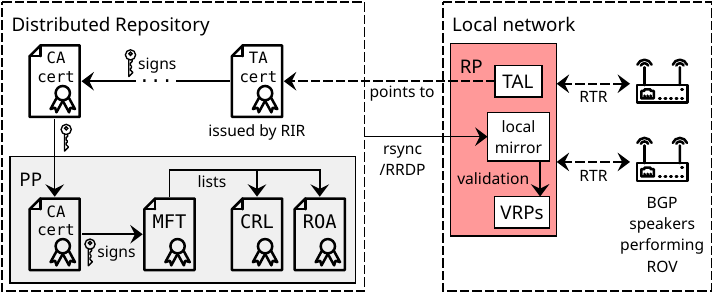}
    \vspace{-5pt}
    \caption{Simplified overview of \acs{rpki} operations.}
    \label{fig:rpki}
    \vspace{-\baselineskip}
\end{figure}

Routing information exchanged by \acs{bgp} speakers does not carry any verifiable authenticity.
This makes \acs{bgp} susceptible to attacks, such as prefix origin hijacks, where malicious actors claim origin for an IP prefix and redirect traffic to their \ac{as}.
\ac{rpki} was designed to add security to inter-domain routing with BGP, making use of the principles of Public Key Infrastructure to bind the ownership of Internet resources, e.g., IP addresses or \ac{as} identifiers, to the ownership of a public key.
Figure~\ref{fig:rpki} gives an overview of \ac{rpki}.
Resource owners act as \acp{ca} owning a certificate that binds their key to their resources.
At the root of the \ac{rpki} hierarchy are the \acp{rir} serving as trust anchors.
They authenticate subordinate \acp{ca}, e.g., \aclp{lir}, which can have their own children.
\Acp{ca} issue \ac{rpki} objects to make statements on the valid use of their resources in BGP.
Prominently, \acp{roa} allow a certain AS to announce a set of IP prefixes in BGP.
Other objects include \acp{crl} for revoking objects, and manifests containing signed lists of all objects issued by \iac{ca}.
Objects are stored on RPKI repositories hosted as \acp{pp}.

To validate BGP announcements with \acp{roa}, networks need to set up a \acf{rp} validator.
\Iac{rp} fetches and validates all objects from all available RPKI repositories and maintains a local cache.
Periodically, the \acp{rp} synchronize their caches with the \acp{pp}, using either rsync or the \acf{rrdp}.
While rsync mirrors a file system directory, \ac{rrdp} servers host HTTPS-accessible notification files listing deltas (incremental changes) and snapshots (complete states) of the repository.
The \ac{rp} parses and validates all \ac{rpki} objects and outputs \acp{vrp}.
\Ac{bgp} speaking routers obtain these from the local \ac{rp} via the \ac{rtr} and perform \acf{rov}, comparing the \ac{bgp} update messages against the \acp{vrp} to detect bogus announcements.

\begin{figure*}[t!]
    \centering
    \includegraphics[width=\linewidth]{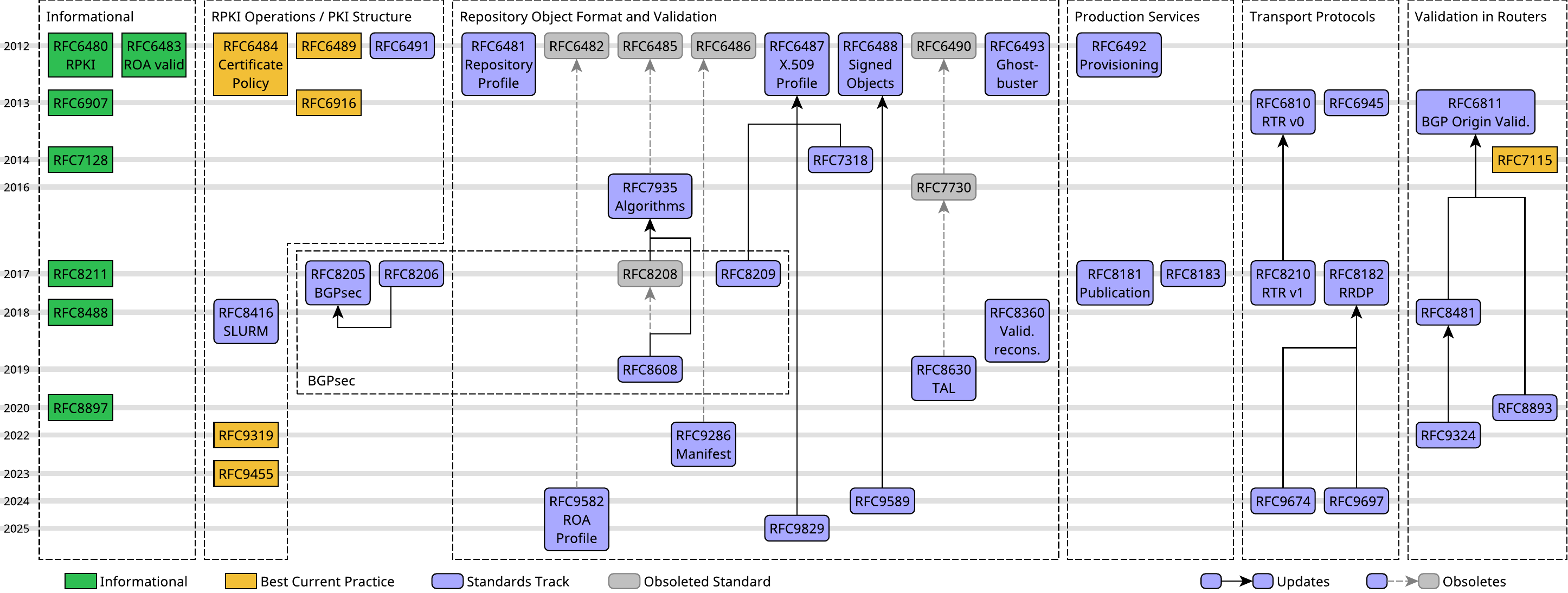}
    \caption{Evolution of RPKI specification in RFCs.}
    \label{fig:rpki-rfcs}
    \vspace{-\baselineskip}
\end{figure*}

\Ac{rpki} is specified in \ac{rfc} documents.
Each RFC defines a specific component of the architecture and generally contains a mix of implicit and explicit requirements for implementations and stakeholders.
Explicit normative requirements make use of well-defined BCP14 upper-case words MUST/SHOULD/MAY etc. to unambiguously express if and how requirements are binding.
RFCs build upon, clarify, update, or partially or fully replace another.
The core components of \ac{rpki} are specified in 50 RFCs, starting with \rfc{6480} published in 2012.
We illustrate this in Figure~\ref{fig:rpki-rfcs}.
Informational RFCs provide general, non-binding recommendations to operators and implementers.
Seven RFCs cover the general operations, PKI structure, and repository organization, including the \rfc{6484} \ac{cp} defining the use and life-cycle of certificates and \rfc{6481} repository profile.
The majority of normative RFCs define the \ac{rpki} object syntax, semantics, and validation logic, like \rfc{6492, 9455, 9582} on \acp{roa}.
Non-certificate objects are embedded as content in signed objects (\rfc{6488, 9589}). 
Certificates follow the RPKI's PKIX profile (\rfc{6487, 7318, 9829}), building upon the X.509 specification's \rfc{5280} general profile and \rfc{3779} IpAddrBlocks extension.
Protocols for production services, specifying the management of repository contents, are drafted in \rfc{6492, 8181}.
\Ac{rrdp} is mainly specified in \rfc{8182}.
A detailed breakdown of all \ac{rpki} RFCs is in Appendix~\ref{app:rpki-rfcs}.

According to NLNetlabs \ac{rov}-measurement~\cite{rovmeasurement}, the currently most popular \ac{rp} implementations are Routinator (72\%), rpki-client (21\%), and FORT-validator (5\%).
We include these three \acp{rp} in our study.
A small fraction ($<3\%$) of other \acp{rp} exists, such as OctoRPKI and RPKI Validator, which are no longer maintained and are thus excluded from our study.

\section{Related Work}\label{sc:works}

{\bf Vulnerabilities in RPKI.} 
Previous research discovered vulnerabilities in RPKI~\cite{gilad2017we,hlavacek2022stalloris,van2022rpkiller,mirdita2023cure,mirdita2024sok,cattepoel2024posterkrill,jacobsen2024posterfort, schulmann2024rpki}, with most issues related to parsing logic. 
These were typically attributed to developer errors or implementation bugs rather than to problems in the underlying specifications. 
Our research traces the root causes of the bugs to flaws in the RPKI specification, showing that a substantial portion of the issues originates from ambiguities, contradictions, or under-specified behavior in the standard documents themselves.

{\bf Implementation bugs.} 
Research has utilized insights into the underlying protocols to improve vulnerability detection in implementations~\cite{pham2020aflnet, tian2019differential, meng2024large, zou2021tcp}, without tracing the issues to the factors causing them, such as implementation errors or flaws in the specification. 
We use implementation issues as a vector to identify problematic sections of the specification that lead to faulty implementation. 
For example, vague or conflicting requirements lead to divergent implementations, even in the absence of implementation errors. 
Such issues cannot be identified by the techniques used in previous work, as those treat RFCs as the ground truth of the protocol rather than as a potential source of implementation issues.

{\bf Formal analysis of specifications.}
Detecting issues in the specification differs from identifying implementation issues and has historically been difficult: The specification is complex, the RFCs have inter-dependencies through references, updates, or obsoletions, and subtle issues in packet formats, state transitions, or corner cases may only appear during execution in a real-world implementation. 
Consequently, flawed or underspecified RFC recommendations, which potentially expose systems to detrimental attacks, often remain undiscovered in the specification for decades. 

A common approach in research for identifying problems in RFCs is based on the formal verification of protocols.
First, the protocol behavior is extracted from RFCs and translated into a formal model using a specification language.
Then, proof techniques or automated verification tools are used to identify logical or security flaws in the protocol described by the RFC. 
Formal verification focuses on self-contained, closed systems; this is not suitable for the analysis of RPKI, which lacks a unified model of the entire system and operates in a distributed and asynchronous environment. 
Formal verification can reason about specification correctness in a standard when applied to an individual component, e.g., handshake logic, \cite{woodage2019analysis,okumura2020formal,yen2021sage,pacheco2022automated}. 
For instance, formal verification analysis of specific security properties of the deterministic random bit generator in the NIST SP 800-90A standard showed that the design was vulnerable~\cite{woodage2019analysis}. 
The authors concluded that their attack was theoretical, demonstrating that the standard specification did not achieve weak forward security. 
A formal analysis of the Web Payments API (WPA) found two vulnerabilities in the specification~\cite{do2022formal}.
The authors did not demonstrate a practical impact but suggested that WPA were expected to be adopted by payment providers in the future.

However, formal verification has inherent key limitations. 
A fundamental issue is that the construction of a formal model is prone to errors. 
This is further exacerbated when the specification is complex, and RFCs contain inter-dependencies with references and updates that obsolete older RFCs. 
As a result, vulnerabilities stemming from underspecification or vagueness may not even be expressible in a formal model. 
Indeed, although recent research showed DNSSEC to be vulnerable to downgrade attacks~\cite{DBLP:conf/uss/HeftrigSW23}, an earlier formal analysis of the DNSSEC standard proved the protocol's correctness and security~\cite{bau2010security}. 

Another inherent limitation is that formal verification tools suffer from state explosion, which makes them unsuitable for fully analyzing protocols that consist of multiple components, such as DNSSEC and RPKI.
Therefore, previous studies encompass a single mechanism or an isolated component that corresponds to one RFC. 
Our methodology enables us to perform the first comprehensive study of a complex mechanism composed of multiple components (trust anchors, RPKI objects -- certificates, \acp{roa}, manifests, \acp{crl} -- \acp{pp}, \acp{rp}, \ac{rov}, transport protocols, ...) which correspond to 50 RFC documents on RPKI.
The inter-RFC dependencies in RPKI, which reference, update, or obsolete others, as well as non-obvious interactions, make it hard to find what to verify.
The effort to build a comprehensive formal model is huge, error-prone, and leads to state explosion.\\
\indent Finally, formal verification typically applies to abstract models, proving the correctness of a simplified or ideal model, which does not capture the impact in the real world, including implementation-specific behavior or deployment-specific assumptions (e.g., NAT devices or firewalls); hence, it overlooks aspects that are implementation- or operation-specific. 
Fuzzing captures realistic models, e.g., \ac{pp} unresponsiveness and how \acp{rp} react to this, which cannot be directly formalized using formal methods.
Classifying which vague formulations constitute a problem with practical impact is challenging based solely on specification text.\\
\indent In contrast to the analysis of specifications using formal methods, we develop a systematic approach based on real-world impacts derived from the analysis of scientific work, GitHub Issues, and CVEs, as well as differential fuzzing of the implementations and evaluations of real-world operations to identify problems in the RPKI specification.
This impact-driven analysis of the specification in the RFCs 
allows us to reveal practical flaws and their real-world impact, which the formal verification models cannot do, since formal methods operate at a mathematical model-based level, abstracted from the implementation.\\ 
\indent {\bf We focus on standard compliant behavior.} 
In contrast to research that explores deviations from the standard, like~\cite{hebrok2023we}, which found that no library implementing the TLS session ticket mechanism fully adheres to the standard recommendations in \rfc{5077}, our goal is to identify flaws in the specification of RFCs. 
We focus on standard-compliant behavior that produces unexpected real-world outcomes, such as failures or vulnerabilities. 
The flaws can be conflicts between different RFCs, vagueness (e.g., due to underspecification), or even vulnerable or buggy recommendations. 

\section{Methodology}\label{sc:methodology}

\begin{table*}[ht!]
    \centering
    \begin{tabular}{l||l|l|l|l|l}
        Name & Sourced & Nr. Issues before & Nr. Issues after & Overlap & RFC-related \\\hline
        Scientific Work & 22 surveyed works & 35 & 35 & 11 (10 CVE, 1 Fuzz.) & 23 \\
        CVE             & 40 CVEs           & 40 & 32 & 13 (4 GH, 9 Sc.w.) & 12 \\
        GitHub          & 601 GitHub issues & 66 & 49 &  5 (4 CVE, 1 Fuzz. \& Sc.w.) & 23 \\
        Fuzzing         & Fuzzing 3 RPs     & 61 & 61 &  1 (1 GH \& Sc.w.) & 23 \\
        Crawler         & 99 repositories   &  3 &  3 &  0 &  3 \\
    \end{tabular}
    \caption{Sources, Nr. of issues before/after de-duplication, Overlap.}
    \vspace{-10pt}
    \label{tab:dataset}
\end{table*}

Our methodology is designed to uncover flaws in RPKI specifications by analyzing how they manifest in validator implementations and operational deployments.
To achieve this, we combine three techniques: (1) differential fuzzing of major RPKI relying party software, (2) large-scale Internet crawling of RPKI repositories to capture real-world deployment behavior, and (3) manual semantic tracing of observed discrepancies to the relevant RFCs. 
Importantly, we do not solely rely on newly discovered issues.
Instead, we combine these results with publicly available information, including CVEs, previous academic publications, and bug reports from open-source repositories. 
This allows us to construct a comprehensive dataset of \emph{impact issues}: practical \ac{rpki} issues that have a security impact by affecting validation outcomes or the observable system state.
In the subsequent RFC analysis, we trace whether these issues are rooted in the RFCs or stem from non-RFC causes, like benign implementer errors.
When an issue is RFC-related, we can deduce that its causes in the RFCs are flaws by construction, as we do not start the analysis from inconsequential specification ambiguities or implementation differences in general, but from confirmed security issues.
 
\begin{figure}[t!]
    \centering
    \includegraphics[width=\linewidth]{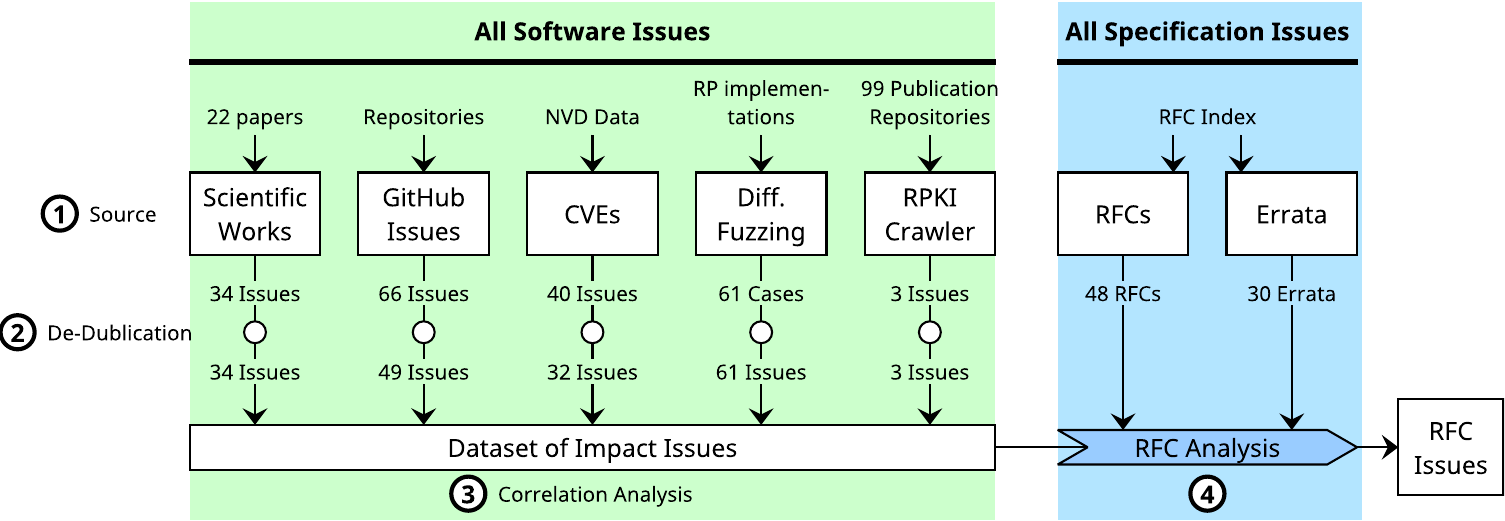}
    \caption{Methodology of the RFC analysis.}
    \vspace{-10pt}
    \label{fig:methodology}
\end{figure}

For each confirmed RFC issue, we generate targeted test cases and analyze validator behavior to assess divergence, compliance, and potential security impacts. 
Our methodology consists of four phases, illustrated in Figure~\ref{fig:methodology} and quantified in Table~\ref{tab:dataset}.

\textbf{\circled{1}~Data aggregation.}
We collect a dataset on problems in \ac{rpki} from past reports and using automated testing.
We source existing reports on problems in \ac{rpki} implementations from scientific publications, CVEs, and GitHub Issues.
To identify new issues in \ac{rpki}, we run a fuzzing campaign based on the publicly available fuzzer CURE~\cite{mirdita2023cure} to study \ac{rp} implementations and develop a crawler to find issues in currently deployed \acp{pp}.

\textbf{\circled{2}~De-duplication.}
To quantify the output of each source, we first perform de-duplication on the extracted data.
We manually group reports on issues if they have equal underlying causes, e.g., two CVEs for distinct implementations relating to the same vulnerability. 
This is based on descriptions/comments of CVEs, GitHub-Issues, scientific publications, and publication timing.
Documentation quality varies, so this is done case-by-case.

\textbf{\circled{3}~Correlation analysis.}
We compare how the sets of issues yielded through each of our source channels contribute to the final dataset and quantify the overlap.

\textbf{\circled{4}~RFC analysis}
Based on the resulting dataset, we conduct an analysis of the \ac{rpki} RFCs.
We systematically map the issues to the clauses in the relevant RFCs.
We analyze the causal relationship between problems in specification and how these problems translate into implementation and deployment.
Based on our results, we systematize patterns of how issues in the \ac{rpki} RFCs manifest and identify classes of problems.

\subsection{Data Aggregation and De-duplication}
Many issues in \ac{rpki} have been identified in the past.
These range from vulnerabilities in implementations to operational problems and inconsistent validation.
We systematically collect issues from these sources to create a dataset of \ac{rpki}-specific issues:

\textbf{Scientific Publications.}
We include all issues from RPKI publications from the past 5 years that mention the terms \textit{vulnerability, issue, inconsistency, or attack}. 
We exclude problems that are the result of \ac{rpki} interacting with other protocols and infrastructures, such as DNS components interacting with \ac{rpki}~\cite{hlavacek2022behind} and discussions of \ac{rpki} design.
From 22 surveyed works, we identify 35 issues in seven publications~\cite{van2022rpkiller, hlavacek2022stalloris, mirdita2023cure, jacobsen2024posterfort, mirdita2022poster, cattepoel2024posterkrill, hlavacek2023beyond} that correspond to our criteria.

\textbf{CVEs.}
We utilize the keyword search of the National Vulnerability Database (\url{https://nvd.nist.gov/vuln}) using the keywords \textit{RPKI}, \textit{RTR}, \textit{RRDP}, and names of the most popular \ac{rp} implementations (\textit{Routinator}, \textit{Fort}, \textit{rpki-client}).
In total, we find 40 CVEs.
After coalescing CVEs that describe the same problems across implementations, we obtain 32 issues.

\textbf{GitHub Issues.}
Additionally, we analyze the GitHub Issues for the repositories of the three \ac{rp} implementations.
At the time of writing, there are 601 GitHub Issues.
We filter out any unconfirmed issues, feature requests, and problems not impacting RPKI operations (e.g., bugs in logging), as well as issues related to third-party software, leaving 66 GitHub Issues.
We group those with equal causes, whether within or between implementations, into 49 issues to be added to our dataset.
We also leverage GitHub's cross-linking: if multiple issues are linked in comments and fixed through a shared commit/pull-request, we de-duplicate them (e.g., Fort GitHub-Issues~58--60 are expressions of one underlying cause).

The analysis of contextual \ac{rpki} publications and reports yields past problems; most have already been addressed in current implementations and deployments.
To systematically identify security issues that affect current systems, we additionally perform new studies on \ac{rp} implementations and \ac{pp} deployments using fuzzing and a crawler.

\indent \textbf{Fuzzing campaign.}

RPKI is a global trust infrastructure whose purpose is to produce consistent validation outcomes across independently deployed validators.
A fundamental requirement for such a system is determinism: given the same repository state, all standard-compliant validators must derive the same validation result.
We conduct a structure-aware differential fuzzing campaign on \ac{rpki} \ac{rp} implementations to identify current inconsistencies in their processing, i.e., to find inputs that lead to divergent validation results.
If validators produce different results on the same input, this violates determinism and leads to inconsistent trust decisions across the ecosystem.
This breaks the guarantee of uniform policy enforcement, i.e., that the same routing object is treated consistently by all the RPs.
For this reason validation inconsistency is inherently security critical, it undermines the correctness and reliability of the trust infrastructure itself, even if a concrete exploit is not immediately demonstrated.
The violation of determinism alone already constitutes a security weakness in a system whose purpose is to provide globally consistent trust decisions.
Validation inconsistencies are thus interesting for RFC analysis: if they result from misinterpretation or problems in the specification, validation inconsistencies in RPs uncover systemic problems in the RPKI RFCs.
Therefore, we complement the dataset of problems with validation inconsistencies detected through fuzzing.

For our analysis, we extend the CURE fuzzer~\cite{mirdita2023cure} to support inconsistency detection and de-duplication. 
We run it 3x1h on each RPKI object type, totaling around 20 million test-objects. 
We use structure-aware mutations guided by the RPKI object grammar and encoding rules \cite{pham2019smart}. 
This ensures that generated objects are parseable and reach the semantic validation logic, where RFC inconsistencies and ambiguities occur, maximizing the number of detected issues. 
An interested reader is referred to Appendix~\ref{app:fuzzer} for more details. 

Using our fuzzer, we find 61 unique inconsistencies between the different RP implementations, identified through differences in the validated output of RPs and their error logs.
38 inconsistencies (62\%) stem from certificate processing. 
This relatively large share is expected, as certificates are the most complex RPKI objects, with specification mandating 29 different fields and attributes per certificate (e.g., compared to 4 different fields in \ac{roa} content). 
We identify 11 more inconsistencies within the parsing logic, i.e., how the \ac{rp} handles structural elements, illustrating that inconsistencies stem not only from object processing but also from basic parsing functionality.
Additionally, we find 12 inconsistencies that occur within non-certificate object validation: \ac{roa}~(5), \ac{crl}~(6), and manifest~(1). During our fuzzing campaign, we also identified two vulnerabilities causing crashes: one in Fort and one in Routinator. 
We disclosed both vulnerabilities to the vendors and received CVEs. 
Vulnerability descriptions are in Appendix~\ref{app:vuln}.

\indent \textbf{\ac{rpki} crawler.} 
To assess the real-world impact of RFC specification issues, we developed a custom RPKI crawler that systematically fetches and validates the global set of published RPKI objects and compares the behavior of \ac{rrdp}-repositories to RFC requirements.
We analyze the repository connectivity, transmitted data, and HTTPS behavior. 
Our crawler implements both RRDP and rsync fetching mechanisms and mirrors the repository structures of all five \acp{rir} and the delegated repositories.

In total, we process over 400K files (ROAs, CRLs, manifests, certificates), reflecting the full operational state of the RPKI ecosystem.
The design of our crawler takes into consideration the highly decentralized and latency-sensitive constraints of the RPKI ecosystem.
Our crawler employs a distributed and throttled scheduler to balance coverage, speed, and ethics, e.g., by adhering to \ac{rrdp} minimum repeat query intervals for \acp{rp}.
It integrates with the differential analysis of RP implementations in a controlled environment to capture validation divergence through validator logs, outcomes, and error codes.
The core engineering innovation of our crawler lies in the provenance tracking engine, which constructs a dependency graph linking validation failures to their root causes, e.g., object fields, chain anchors, manifest mismatches, or deployment issues.
This allows us to differentiate between failures due to malformed objects and those stemming from under-specified or contradictory RFC semantics.

This component allows us to bridge the gap between abstract RFC analysis and its concrete operational consequences. 
By quantifying how many objects are affected by each class of flaw discovered through fuzzing or manual analysis, we can systematically analyze their prevalence and real-world relevance.
Moreover, we utilize the crawling mechanics to measure divergence in \ac{rrdp} behavior of \acp{pp}, including message payloads, transport metadata, and availability behavior.
Through this, we identify 3 issues leading to inconsistent behavior of \acp{pp}: availability, \ac{rrdp} file syntax, and TLS certificate deployment. 
An interested reader is referred to Appendix~\ref{app:crawler} for a more detailed description of the crawler implementation.

We open-source the full collected dataset and our implementations as artifacts.\footnote{\url{https://github.com/tkh42/rfc-analysis}}

\subsection{Correlation Analysis}
We evaluate how each data source contributes to the final dataset by determining the overlap between the collected issues, i.e., which \ac{rpki} issues we identify in more than one source, listed in Table~\ref{tab:dataset}.
We find that 11 issues (31\%) of scientific publications overlap with issues from CVEs (10) and GitHub Issues/Fuzzing (1).
5 issues (10\%) from GitHub Issues are also discussed in CVE issues (4) or scientific works/fuzzing (1).
The minor overlap between GitHub issues and scientific work shows a discrepancy in how identified problems are reported in \ac{rpki}, with researchers not opening issues on GitHub and contributors to GitHub not writing scientific publications. 
Inversely, 13 issues (41\%) from CVEs are also retrievable from either scientific works or GitHub Issues.
We observe only one overlapping issue found with fuzzing and no overlap for the crawler issues, which shows that our practical studies almost exclusively identified new problems.

Generally, none of our data sources is fully encompassed by the others, and the overlap between all sources is small, illustrating that utilizing different sources to obtain a comprehensive overview of practical \ac{rpki} issues is beneficial.

\subsection{RFC Analysis}
With the dataset of issues, we systematically investigate the root causes in the specification RFCs.
To aid this process, we first map the \ac{rpki} RFCs by purpose in the RPKI infrastructure, as visualized in Figure~\ref{fig:rpki-rfcs}.
Thus, when an issue relates to the implementation of a specific part of the infrastructure, e.g., publication, we narrow the set of relevant RFCs for analysis.
We incorporate the RFC errata in this process, as they provide insights into previously identified problems with the specification text.

For each dataset issue, we perform a manual analysis consisting of the following steps:\\
1. We collect all relevant RFC sections. This is aided by performing a keyword search over the RFC text to identify sections that share terminology with the issue.\\
2. We investigate how the phrasing of the specification relates to the issue.\\
3. We classify whether the issue has roots in the specification. 
We document novel findings on how specification problems translate into implementation and deployment.\\
4. We evaluate the operational impact of the issue by developing implementation-specific test cases.

RFCs may allow flexibility in how implementations realize certain behaviors, including differences, such as in processing, data structures, or error handling.
Such flexibility is explicitly defined in the specification and operates at the level of implementation choices.
However, specification flexibility is bounded by its security impact.
In a global trust infrastructure, such as RPKI, acceptable flexibility must not affect security relevant properties, in particular the consistency of validation results across independently deployed validators.
We therefore define specification flexibility as permissible only as long as it preserves security-relevant equivalence of system behavior.
In contrast, the RFC issues we identify lead to observable negative security impact, such as divergent validation outcomes, VRPs or protocol behavior resulting in inconsistent trust decisions.
This security impact indicates that the specification fails to define acceptable behavior in a way that preserves security-relevant equivalence across implementations and we therefore classify it as a specification flaw rather than an expression of flexibility.
We therefore start from impact issues as indicators of candidate specification flaws.
This allows us to distinguish intended flexibility from under-specification, vagueness or conflicting requirements that affect trust decisions.
Our analysis classifies such specification flaws into the following categories, capturing how they arise in RFCs:

\textbf{Conflicting requirements.}
RFCs frequently contain multiple distinct requirements that apply to the same functionality.
For instance, in RPKI, whether an object is considered `valid' is subject to several distinct requirement conditions.
When these requirements are in opposition, implementations cannot fully comply with the specification and must make deliberate choices on which parts of the specification take precedence. 
For example, implementing validation of PKIX certificate extensions, \rfc{6487} states that the extensions not explicitly mentioned in §1 MUST be absent, but also allows ignoring unknown extensions. 
Identifying and contextualizing all relevant, overlapping requirements can reveal such conflicts and allow us to identify how they lead to inconsistent implementation. 

\textbf{Unspecified cases.}
Specifications need to be precise and deterministic for any conditional decisions, such as handling inputs or validation states, and instruct behavior in corner cases.
When specification language is absent from RFCs, implementations need to make their own assumptions on how to handle these unspecified cases.
This can lead to inconsistency and vulnerabilities.
For example, the RRDP RFC does not define how processing should proceed after a failed hash verification of a file in a notification, requiring \acp{rp} to decide whether to ignore this single file or abort the entire update operation.  
We classify RFC issues based on whether they lack explicit requirements for all possible cases.

\textbf{Vagueness.}
Vagueness is imprecise language where specification is formally complete (all cases are covered) but allows for multiple diverging interpretations due to phrasing.
This forces implementers to concretize the requirements themselves, leading to inconsistencies.
This is inherently different from legitimate RFC flexibility where RFCs permit explicit divergence on \textit{how} things are implemented, but which should not lead to security impact such as different validation results on the same object.
We classify vagueness based on whether two different, conflicting implementations are possible and valid according to a given requirement.

\textbf{Complexity.}
The nature of RFCs leads to related requirements being distributed over a large number of distinct RFC documents.
Even if the RFC language is precise, this can lead to operational issues when implementers are unaware of the full context of a requirement while implementing specific RFC requirements.
We define complexity as cases in which interconnected requirements for a feature are distributed among multiple RFCs or multiple sections within an RFC without adequate, concrete references.
For example, \ac{rpki} certificate extension processing steps are defined in \rfc{6487} as differences to the validation of the general PKIX \rfc{5280}, selectively and partially overriding requirements without contextualization.

\textbf{Infeasible requirements.} 
RFC requirements must be translated into real-world deployments and can prove operationally infeasible; e.g., if the requirement leads to validation failures in benignly misconfigured objects.
This manifests in implementations that deliberately oppose RFC specification, traceable through documentation on issues and inconsistent behavior.

\section{Flaws in RPKI Standard Specification}
\label{sc:rfc-analysis}

\newcommand{\findingrcl}{Repository Chain Limit}
\newcommand{\findingnfh}{Notification File Hash}
\newcommand{\findingrrdpdup}{RRDP Entry Duplication}
\newcommand{\findingfrfc}{Formality in \rfc{6487}}
\newcommand{\findingasnint}{ASN.1 INTEGER}
\newcommand{\findingseravail}{Service Availability}
\newcommand{\findingtermdef}{Term Definitions}
\newcommand{\findinghttpscert}{HTTPS Certificate Validation}
\newcommand{\findingderenc}{Use of DER Encoding}
\newcommand{\findingipaddrrepr}{IP Address Representation}
\newcommand{\findingcpqvalid}{\texorpdfstring{CP Qualifiers and \rfc{8360}}{CP Qualifiers and RFC {8360}}}
\newcommand{\findingextensions}{X.509 Extensions}

\newcommand{\C}{\CIRCLE}

\begin{table*}[t]
    \centering
    \footnotesize
    \renewcommand{\arraystretch}{0.8}
    \begin{tabular}{llc"w{c}{1em}|w{c}{1em}|w{c}{1em}|w{c}{1em}|w{c}{1em}"w{c}{1em}|w{c}{1em}|w{c}{1em}|w{c}{1em}|w{c}{1em}"w{c}{1em}|w{c}{1em}|w{c}{1em}|w{c}{1em}"w{c}{1em}|w{c}{1em};}
        \multicolumn{3}{l"}{Issue} &  \multicolumn{5}{@{~}c@{~}"}{Sources Used} & \multicolumn{5}{c"}{Categorization} & \multicolumn{6}{c;}{Operational Impact} \\
        & & & \multicolumn{5}{c"}{} & \multicolumn{5}{c"}{} & \multicolumn{4}{c"}{\acs{rp}} & \multicolumn{2}{c;}{Repos.} \\
        & & Novel & SW & CVE & GH & DF & RC & V & CR & UC & C & IR & NC & IV & MF & DoS & NC & OR \\\hline\hline
        
        1  & \findingcpqvalid{}   & $\ast$ &    &    &    & \C &    &    & \C & \C &    &    &    &    & \C &    &    &    \\\hline  
        2  & \findingextensions{} &        &    &    & \C & \C &    & \C & \C & \C & \C & \C & \C & \C &    &    &    &    \\\hline  
        3  & \findinghttpscert{}  & $\ast$ &    & \C & \C & \C & \C &    &    &    & \C & \C & \C &    & \C &    &    & \C \\\hline 
        4  & \findingderenc{}     & $\ast$ &    &    & \C & \C &    & \C &    &    &    & \C & \C & \C &    &    & \C &    \\\hline 
        5  & RRDP Withdraw        & $\ast$ &    &    &    & \C &    & \C &    & \C &    &    &    & \C &    &    &    &    \\\hline 
        6  & \findingtermdef{}    &        & \C &    & \C & \C &    & \C &    &    & \C &    & \C & \C &    &    &    &    \\\hline 
        7  & \findingrcl{}        &        & \C & \C &    &    &    & \C &    & \C &    &    & \C &    & \C & \C &    &    \\\hline 
        8  & \findingseravail{}   &        & \C & \C & \C &    & \C &    &    & \C &    & \C &    &    &    & \C & \C &    \\\hline 
        9  & \findingasnint{}     & $\ast$ &    &    & \C & \C &    & \C &    & \C & \C &    &    & \C &    &    &    &    \\\hline 
        10 & \findingipaddrrepr{} & $\ast$ & \C &    &    & \C &    &    & \C &    & \C &    & \C & \C &    &    &    &    \\\hline
        11 & \findingrrdpdup{}    & $\ast$ &    &    &    & \C &    & \C &    & \C &    &    & \C & \C &    &    &    &    \\\hline
        12 & \findingnfh{}        & $\ast$ &    &    &    & \C &    &    &    &    & \C &    & \C & \C &    &    &    &    \\\hline 
    \end{tabular}\\[.2pt]
    {\footnotesize \textbf{Sources.} SW: Scientific Works, GH: GitHub Issues, DF: Differential Fuzzing, RC: \ac{rpki} Crawler\\[-1.3pt] 
    \textbf{RFC Issue.} V: Vagueness, CR: Conflicting Requirements, UC: Unspecified Cases, C: Complexity, IR: Infeasible Requirement\\[-1.3pt]
    \textbf{Operational Impact.} NC: Non-Compliance, IV: Inconsistent Validation, MF: Missing Feature, DoS: Denial of Service, OR: Objects rejected}
    \caption{
    \centering
    Identified issues, the methods we used, their categorization and operational impact in \acp{rp} and repositories.
    }
    \label{tab:rfc_issues}
    \vspace{-1.5\baselineskip}
\end{table*}

Applying our methodology, we identify 12 problems within the \ac{rpki} RFCs with practical impact on \ac{rpki} implementations and deployments.
An overview is given in Table~\ref{tab:rfc_issues}, also listing which identified issues are novel.
Four RFC problems (2,\,6,\,7,\,8) have been reported on previously, either as errata or in scientific contributions; however, our analysis offers new insights by identifying the underlying RFC sections and evaluating their impact.
The other eight RFC problems are novel.
For each identified flaw, we create test cases to evaluate how the requirement is implemented, detecting non-compliance, inconsistent validation, missing features, \ac{dos} vectors in \acp{rp} and non-compliance in \acp{pp}. 
We additionally test whether they cause the rejection of real-world standard-compliant \ac{rpki} objects. We discuss the identified flaws next. 

\subsection{\findingcpqvalid{}}
\label{sc:cpqvalid}

\begin{figure}[t!]
    \centering
    \includegraphics[width=\linewidth]{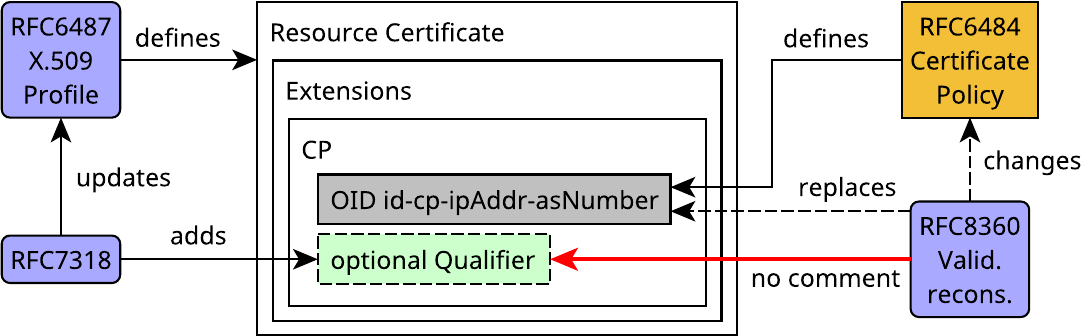}
    \caption{Missing comment on \acs{cp} Qualifiers in \rfc{8360}.}
    \vspace{-5pt}
    \label{fig:cp-valid}
    \vspace{-\baselineskip}
\end{figure}

\rfc{7318} (Jul 2014) updates \rfc{6487}[4.8.9] (Feb 2012) by allowing the optional inclusion of zero or one \ac{cp} qualifier in the \aclp{cp} extension.
Moreover, \S3~acknowledges that strictly following old \rfc{6487} validation invalidates \rfc{7318}-conforming objects.\\
\indent In \rfc{8360} (April 2018), an alternative profile for PKIX Resource Certificates was published. 
This update changes the object validation algorithm to be more error-resistant without changing the syntax and, to avoid inconsistent implementations, defines new OIDs to signal the new algorithm.
For this, it updates \rfc{6487}, but conflictingly replaces \rfc{6487}[4.8.9] again, not acknowledging the \rfc{7318} changes.
We illustrate this conflict in Figure~\ref{fig:cp-valid}.\\ 
\indent \textbf{Implementation compliance.} 
We test the support of \rfc{8360} validation by encoding a certificate and an entire publication point with the new OIDs.
We find multiple inconsistencies.
First, rpki-client does not support the new OIDs and rejects any object using them, while Routinator and Fort support them and apply the new validation algorithm. 
Second, we find that Fort only succeeds validation if both the \ac{roa} and its respective certificate use a consistent set of OIDs, which is neither mandated in the RFC nor necessary for security. 
Investigating the Fort code, we find this constitutes a bug in the validation procedure failing to handle mixed \ac{cp} validation. \\
\indent The observation of these inconsistencies puts the idea of the new validation policy into question.
It was designed to ensure child certificates remain valid if unrelated parent resources change and the change is not fully propagated to the children. 
However, with the lacking or faulty support in rpki-client and Fort, using the new policy renders the certificate invalid, also recursively invalidating all its child repositories. 
If, e.g., RIPE NCC were to issue a new certificate with the \rfc{8360} policy to supposedly increase resilience against errors, it would instead invalidate the entire RIPE NCC RPKI tree in any system using rpki-client or Fort.

\subsection{\findingextensions{}}
\label{sc:extensions}

Within the specification of X.509 certificate extensions, we identify a multitude of findings.
\rfc{6487}[1] specifies that \textit{extensions not explicitly mentioned MUST be absent} in resource certificates.
However, later in §4.8, the formal requirements are:\\
1. The extensions listed in §4.8 ``MUST be present in a conforming resource certificate, except where explicitly noted otherwise.''\\
2. Systems ``MUST reject the certificate if it encounters a critical extension it does not recognize.''\\
3. Systems ``MAY ignore unrecognized non-critical extensions.''\\
\indent While these three requirements do not contradict each other, point~(3), allowing that unrecognized extensions ``MAY'' be ignored, directly contradicts \S1.
Moreover, this also contradicts the more recent informational, hence explicitly non-normative, \rfc{8897}[3.1] on requirements for \acp{rp} stating ``any extension excluded by Section 4.8 of [RFC6487] must be omitted.''
This conflict has also been mentioned in erratum \textit{eid3168} but persists to date.
The conflicting requirements on how extensions must be validated by \acp{rp} leads to a slate of vague specifications and undefined cases.

Complicating this issue, \rfc{6487} explicitly inherits PKIX general specification (\rfc{5280}) and its requirements, without clarifying how inherited requirements interact with new requirements (1), (2), and (3). 
For example, \rfc{5280}[4.2] explicitly says that a certificate ``MUST NOT include more than one instance of a particular extension.''
An \ac{rp} could either ignore a duplicate non-critical, non-\ac{rpki} extension and assume validity as \rfc{6487}[4.8] suggests, or fail validation due to \rfc{5280} requirements.
Cases of invalid, ill-structured, duplicate, or syntactically correct extensions with unrecognized OIDs are not clearly defined.\\
\indent \textbf{Implementation compliance.}
We find many cases of inconsistent processing of X.509 extensions. 
Rpki-client is the only RP that accepts critical extensions that are unknown to RPKI specification but known to OpenSSL, like subject alt name \rfc{4985}.
Code investigations show that rpki-client relies on OpenSSL for initial extension validation. 
Since the extension is known to OpenSSL, the extension is accepted even though rpki-client cannot process it.
Routinator is the only implementation not failing validation if duplicate unknown extensions are present in a certificate. 
While all \acp{rp} accept an unknown non-critical extension if its internal formatting is correct, only rpki-client and Fort fail if the encoded extension value cannot be decoded. 

Evidently, overriding requirements from established technologies, like PKIX, can cause substantial inconsistency when implementations incorporate existing libraries that implement the original requirements, leading to conflicts.

\subsection{\findinghttpscert{}}
\label{sc:stricthttp}

RPKI uses HTTPS to ensure transport security.
However, \ac{rrdp} \rfc{8182} mandates that RP implementations ignore HTTPS validation failures.
Specifically, \acp{rp} must continue to retrieve data even if HTTPS validation fails, as this is usually a symptom of benign configuration errors and RPKI objects are protected by signatures anyway.
This design contradicts the validation procedures in HTTPS \rfc{8446}.

\textbf{Implementation compliance.}
We test \acp{rp} by setting up a valid repository with a self-signed, an expired, a malformed, and no HTTPS certificate.
No current RP implementation follows the \rfc{8182} requirements;
they stop retrieving data if a repository has a faulty HTTPS certificate, in line with how HTTPS was designed.
Interestingly, proper adoption of the RFC requirement in the now deprecated RIPE validator has been scolded in CVE-2020-16163, complaining about the implementation dissenting HTTPS validation rules.
The issue shows that defining requirements that go against the initial use-case of a technology, like HTTPS, leads to non-compliance if the benefit is insufficient to warrant the overwrite.

\subsection{\findingderenc{}}
\label{sc:derenc}

\rfc{6488} (signed objects) mandates that all \ac{rpki} objects must use DER encoding to ensure the encoding is predictable and deterministic.
This imposes stricter rules on RPKI objects than the general CMS specification in \rfc{5652}, which allows more generic encoding schemes like BER and CER, or the X.509 encoding requirements, which mandate DER for ``data to be signed'' (\rfc{5280}[4.1]), but not the entire object.

\textbf{Implementation compliance.}
We create a test case by encoding a valid \ac{rpki} object in CER.
No RP enforces the requirement from \rfc{6488} that objects must be DER encoded.
This is likely rooted in operational necessity.
Manual investigation shows that a subset of otherwise valid CAs (1,627 objects, 0.3\% of objects) use non-DER compliant encoding, impacting 1,952 prefixes.
These would not be available if RPs strictly followed RFC encoding rules.
The non-compliance illustrates implementations will disregard non-essential RFC requirements if they improve practical resilience against misconfigurations.

\subsection{RRDP Withdraw}
RPKI offers two separate mechanisms to remove a non-expired object from a CA to exclude its content from the validated output.
CAs can either add the object to the CRL and revoke its certificate, or withdraw the object through a \textit{withdraw} entry in an \ac{rrdp} delta file.
While revocation with \acp{crl} is comprehensively defined in \rfc{6487}, \ac{rrdp} \rfc{8182} is vague on the implementation of withdraws, not defining concrete steps or requirements on how they should be performed.
This causes conflicts if an object is withdrawn through \ac{rrdp}, signaling deletion, but is still listed in a manifest, thus required for validation.
The RFC does not specify whether the signed manifest entry should override the unsigned RRDP withdraw.

\textbf{Implementation compliance.}
We test RP behavior through three tests:

First, we issue a withdraw to a \ac{roa} while keeping it listed in the manifest, testing how RPs interpret the conflicting information.
Fort deletes the object from its local cache, then fails the validation of the manifest as it cannot find the listed ROA.
Routinator and rpki-client detect the conflict and prioritize the signed manifest, ignoring the withdraw and succeeding manifest validation.
Next, we withdraw and unlist the ROA from the manifest, then re-list it in the next delta's manifest to test if RPs respect withdraws once the object is no longer in the manifest.
All RPs delete the ROA from their local cache and thus fail validation once the ROA is re-listed.

Lastly, we test removing and re-adding the ROA to the manifest without a withdraw.
All RPs keep the ROA in their local cache and succeed validation once it is re-added to the manifest.
The issue shows that overlapping mechanisms in RFCs require clear advice on which take precedent.

\subsection{\findingtermdef{}}
\label{sc:termdef}

RFCs take special care in explicitly defining specific terms to ensure no misinterpretation of requirements.
We still find wordings in RFCs that are not clearly defined and lead to diverging interpretations.
A prominent example of this problem is \rfc{6487}[4], requiring that listed fields must be present in a valid resource certificate; any other \emph{field} must not be present.
Yet, the section does not define the term ``field.''
\rfc{6487}[4] instructs that, if not explicitly overruled in \rfc{6487}, \rfc{5280} specifications apply.
Looking into \rfc{5280}, the term \emph{field} is used to generically describe elements of the ASN.1 encoded structure.
When referring to values within some specific fields, including Issuer, \emph{attributes} is used, but no clear distinction is drawn.
With the requirement that no additional \textit{fields} are allowed in a resource certificate, it is unclear if this rule encompasses what \rfc{5280} calls \textit{attributes}.

In practice, the lack of proper definitions results in vagueness in \rfc{6487}[4.4] on the Issuer field in resource certificates.
The \emph{CommonName} in an Issuer field is required, the \emph{serialNumber} is optional, but there is no clarification on whether any other \emph{attributes} are allowed and how their inclusion affect certificate validity.
As neither \emph{attribute} nor \emph{field} have concrete definitions, the RPKI requirements on \emph{fields} are of no aid to implementers.
One interpretation of \rfc{6487}[4] reads Issuer \emph{attributes} being encompassed by the term \emph{field value}.
Then, \rfc{6487}[4.4] (specifying a \emph{field value}) implicitly forbids other unmentioned \emph{attributes} since \emph{field values} as specified in \rfc{6487} ``MUST be used in conforming resource certificates'' (\rfc{6487}[4]).
However, if \emph{attributes} are distinct from \emph{fields} (as the use of these terms in \rfc{5280} implies), \rfc{6487}{4.4} makes no statement on additional attributes, thus not explicitly overriding \rfc{5280}.
In theory, \rfc{5280}[4.1.2.4] should then apply, which explicitly states that implementations ``MUST be prepared to receive certificates'' with certain Issuer standard attribute types, including the \emph{OrganisationName} attribute.
Note that this is itself vague as to what constitutes \emph{prepared to receive}, leading to ambiguity in if and how this rule would be overruled by \rfc{6487}; however, \rfc{5280} does not forbid valid certificates from using these well-known attributes.

\textbf{Implementation compliance.}
The vagueness and complexity led implementers to different conclusions on allowed attributes in resource certificates.
Fort implementers interpreted that no additional attributes must be present within fields, causing Fort to treat more than 6K Amazon prefixes as not `RPKI protected,' since their RPKI objects contained an \emph{OrganisationName} attribute~\cite{mirdita2023cure}.
Our test shows that rpki-client now also applies this interpretation, rejecting objects with the attribute. Only Routinator accepts objects with an additional attribute.
Evidently, without proper guidance on how a term should be interpreted, implementations will make their own assessment.
RFCs thus must ensure terms are fully defined before introducing requirements that rely on correct understanding of the term.
The conflict is yet to be addressed in the specification.

\subsection{\findingrcl{}}

\acp{rp} interact with untrusted servers when downloading repository data. 
\rfc{6481} acknowledges that allowing repositories to create an arbitrary number of children may lead to long validation chains and thereby stall \acp{rp} through deep delegations.
The RFC addresses this by proposing \acp{rp} to implement a maximum chain length, but not providing guidance on what this length should be.
Further, the RFC does not limit the maximum chain breadth, exposing \acp{rp} to attacks by repositories that have arbitrarily many children~\cite{van2022rpkiller}.
\Ac{rrdp} \rfc{8182} further states an \ac{rp} should have \textit{some kind of bound to the amount of work it is willing to do}, not addressing how this should be applied in practice. Even under benign operation, no concrete recommendations on limits can lead to scalability issues. 
Previous work~\cite{mirdita2024sok} found \acp{rp} struggle to find fitting limits when requesting unresponsive repositories, increasing validation times.

\textbf{Implementation compliance.}

To test the repository chain limit, we test \acp{rp} for the maximum allowed repository chain depth and breadth.
We place each repository on its own domain to simulate a realistic deployment of a stalling attack~\cite{van2022rpkiller,hlavacek2022stalloris}.
Our tests show that all \acp{rp} limit the maximum depth, complying with \rfc{6481}. 
The limits differ across \acp{rp}, with rpki-client enforcing a default max. depth of~11, Fort~30 and Routinator~31.
No RP limits the maximum breadth of child repositories for a given CA. 
Thus, creating repositories with many children stalls all RP implementations.
Clearly, relying on vague statements, like requiring that \textit{some kind of bound} should applied, is insufficient for protecting implementations against attacks.

\subsection{\findingseravail{}}
\label{sc:seravail}
Availability of services is essential for the functionality of RPKI.
The RFCs acknowledge this, with \rfc{6481} stating, the ``publication repository SHOULD be hosted on a highly available service and high-capacity publication platform.''
\rfc{6484} is more stringent, saying CAs or repository operators must not ``intentionally use technical means of limiting read access.'' 
This blanket rule disbarring repository operators from any means of limiting access would not allow blackholing or denying access to attackers, even when subjected to DoS attacks.
Prior work~\cite{hlavacek2022stalloris} reported that operators use rate-limiting to safeguard to their server, which conflicts with \rfc{6484}.
This may not align with real-world circumstances, leading to service degradation if CA or repository reliability fails RFC expectations,
for instance, when large delays are caused by unresponsive repositories, observed in~\cite{mirdita2024sok}.
For provisioning, \rfc{6492} explicitly acknowledges the necessity of flow control in servers and suggests using HTTP status codes.
However, this is unspecified for RRDP in \rfc{8182}, a vagueness that has been investigated in~\cite{van2022rpkiller}.

\textbf{Implementation compliance.}
We replicate the HTTP response code study in~\cite{van2022rpkiller}, and measure \ac{rp} delays when confronted with low response rates or unresponsive repositories.
First, we set up a repository that replies with an HTTP 404 error, indicating resource unavailability. 
All \acp{rp} immediately stop fetching from the repository.
If a repository replies with an error code of temporal unavailability, like 503, we observe that Routinator and rpki-client treat the repository identically to a 404 error message while Fort retries after 4 seconds, subsequently concluding that the resource is unavailable.
This is a change to the results in~\cite{van2022rpkiller}, where no \ac{rp} was honoring response codes.
Additionally, we test if a slow replying repository can stall \acp{rp} by setting up a repository rate-limited to 100 bytes/s.
We find that Fort implements a protection against a slow-replying repository, aborting the connection after 24s if the answer rate is below 10kB/s. 
Routinator and rpki-client get stalled, continuing the slow connection until all data has been downloaded or the RRDP timeout value has been reached: 360s in Routinator and 900s in rpki-client.
While these protections are an improvement over \cite{van2022rpkiller},
they are still insufficient to protect against long stalling attacks, as attackers may abuse long validation chains and stall \acp{rp} at every link. 

\subsection{\findingasnint{}}
\label{sc:asint}
The ASN.1 object specifications in RPKI have two distinct types for AS numbers.
In the \rfc{3779} AS Identifier Delegation extension, the \emph{ASId} type is an ASN.1 INTEGER and used for the min/max of \ac{as} number ranges.
Similarly, in \rfc{6482} ROA eContent, the \emph{ASID} type is an ASN.1 INTEGER with the concretization that the field \emph{asID} of type \emph{ASID} holds \iac{as} number.
Both types are vaguely defined, lacking formal bounds or a reference to what constitutes an AS number.
In BGP, AS numbers are 32-bit representable unsigned integers, but ASN.1 INTEGERS can be negative or larger than 32-bit.
It is unspecified how unexpected values shall be treated by \acp{rp} when parsing these types.
The lack of bounds to ASN.1 INTEGER decoding is not universal within the RPKI RFCs. 
For example, the ROAIPAddress type within the \ac{roa} profile contains the maxLength field of type INTEGER and has clear bounds in \rfc{6482}[3.3], illustrating that bounds are sometimes applied to decoding and indicating unintended vagueness in the ASId/ASID cases.

\textbf{Implementation compliance.}
We test integer processing in the \acp{rp} by creating a \ac{roa} with \iac{as} number 0xFF.
Since INTEGERs are two-complement numbers and the first bit of 0xFF is 1, its value is $-1$ and thus constitutes an illegal \ac{as} number.
Routinator and rpki-client both detect the error and discard the \ac{roa} as invalid. 
Fort, however, implements integer decoding differently, ignoring that integers are two-complement numbers and instead parsing the value as unsigned, corresponding with how AS numbers are defined (\rfc{6793}). 
Fort thus interprets 0xFF as the valid \ac{as} number 255, leading to an inconsistent cache state between \acp{rp}.
This parsing issue illustrates that distinct, yet similarly named terms with similar purposes introduce complexity and lead to discrepancies in implementation.

\subsection{\findingipaddrrepr{}}
\label{sc:ipaddrrepr}

\rfc{3779} imposes a strict syntax for \emph{IpAddrBlocks} in the PKIX IP Address Delegation Extension.
IP prefixes must be ordered by IP address family, then by bit value of the prefix with no overlapping.
In contrast, the original ROA \rfc{6482} posed no equivalent restrictions for its IpAddrBlocks field, which shares its name with the extension type but has a different definition.
Update \rfc{9582} resolves this and encourages CAs to order the IP address prefixes in ROA similarly to \rfc{3779} and to expect \acp{rp} to enforce this.
BCP \rfc{9455}, published before \rfc{9582}, recommends to not including more than one IP address prefix in \acp{roa} at all.
Specification spread across multiple documents and contradictions lead to inconsistent enforcement.

\textbf{Implementation compliance.} 
We create three certificates with non-canonical \rfc{3779} extensions. 
First, we add two neighboring IP prefixes  
that could be combined into one prefix. 
Routinator accepts the object; Fort and rpki-client reject it as both use OpenSSL for validation, which considers the extension value malformed.
Second, if the extension contains two neighboring prefixes that cannot be combined into one prefix, 
the object is still rejected by Fort and rpki-client as both prefixes could instead be one address range, as is RFC-mandated.
Third, if a range is erroneously used instead of a prefix, 
only Fort rejects the object.

The strict rejection of non-canonical IP addresses is deterministic, but also increases the burden on repository operators, who must ensure perfect encoding of IP addresses to get objects accepted by all \acp{rp}.
The inconsistent enforcement complicates identifying issues, as a non-canonically encoded extension may be accepted by one \ac{rp}.

\subsection{\findingrrdpdup{}} 
The \ac{rrdp} \rfc{8182} does not define how \acp{rp} should handle duplicate entries within a delta, e.g., two identical \acp{roa} being added.
It is unclear if \acp{rp} should ignore the duplicate, fail validation of the delta, or ultimately abort the \ac{rrdp} update; implementations diverge on this issue.

\textbf{Implementation compliance.} 
We find \acp{rp} react differently to duplicate entries, with Routinator falling back to snapshot, rpki-client failing the update and reverting to rsync, and Fort ignoring the duplicate and applying the delta.

\subsection{\findingnfh{}} 
\ac{rrdp} update notification files list one snapshot and a series of delta entries; each entry has a URI and a hash value.
\rfc{8182}[3.4.1] describes the notification file processing by \acp{rp}, referring to \S3.5.1.3 for required validation steps.
In case of failure, the notification file must be rejected.
\S3.5.1.3 lists validation rules that ``{MUST be observed};'' one requires each entry's hash to match the hash sum of the file located under the URI.
In case of mismatch, the \ac{rp} ``MUST reject the file,'' but the specification is vague if this constitutes a notification format violation.
The phrase ``{reject the file}'' implies that only the entry delta/snapshot is rejected, not the notification.
This conflicts with the wording in \S3.4.1, i.e., when any validation step (one is hash verification) fails, the notification file must be rejected and \ac{rrdp} cannot be used.

\textbf{Implementation compliance.}
All \acp{rp} validate the hash of the delta file before applying the update. 
In our test case, we break the delta hash, keeping it 256 bit but invalid. 
While all \acp{rp} detect the broken hash,
Routinator and Fort fall back to the snapshot, as required by the specification, but rpki-client aborts the \ac{pp} update and falls back to the previously stored version in the local cache. 
After the update, the \acp{rp} have an inconsistent state: Routinator and Fort apply the new snapshot while rpki-client remains in the old state. 

\subsection{RFC Non-Compliance}
\label{sc:non-compliance}
The above issues can all be traced back to RFC problems.
In this section, we list issues we found, that arise from RFC non-compliance, i.e., issues in implementations disregarding clear requirements in RFCs.
We find 38 cases of inconsistent validation rooted in direct non-compliance with RFCs, which we classify as follows. 

\textbf{Implementation of non-standard features.}
Only rpki-client implements Elliptic Curve Cryptography (ECC), which is yet to be supported by the RPKI RFCs. Routinator and Fort reject objects signed with ECC. 

\textbf{Ignoring non-canonical object contents.}
The RPKI RFCs are generally strict about disallowing unspecified fields within objects.
These requirements are not security critical but ensure objects adhere to the strict RPKI syntax.
In 16 cases, we observe implementations that do not enforce these checks and ignore non-canonical fields.
Examples include rpki-client and Routinator ignoring additional signedObject URIs in the SubjectInformationAccess extension, defying \rfc{6487}, or Routinator accepting additional qualifiers in the \aclp{cp} extension, defying \rfc{7318}.
The implication of this \ac{rp} non-compliance is poor object hygiene; \acp{ca} have less incentive to ensure that no RFC-undesired fields are in their objects if \acp{rp} accept them nevertheless. 

\textbf{Ignoring value constraints.}
In 20 instances, RP implementations do not enforce checks on object values.
For example, only Fort enforces an object's signedObjectURI to match its actual location.
Furthermore, Routinator ignores a mismatch between the issuer name in a child certificate and the subject name of its parent.
In both cases, these checks are irrelevant to the operational correctness of the \ac{rp}.
In the latter case, Fort and rpki-client use the issuer name in processing to map children to parent certificates, thus enforcing the check, while Routinator does not.
While the enforcement of field value constraints may appear superfluous to \acp{rp}, novel features in \ac{rpki} could change this and make checks security-critical.  
This has already been the case for the AS Number Delegation Extension, which had no applications in \ac{rpki} operations until the introduction of ASPAs.
\section{VRPs' Inconsistencies in the Wild}\label{sc:inconsistencies}
Next, we analyze inconsistencies in deployments to identify real-world impact of issues.

\subsection{Study Methodology}
We install the three \acp{rp} locally and execute them simultaneously every 15 minutes, which we found experimentally to be the minimal possible update cycle without overlapping runs.
To study validation failures during operation, we analyze the VRPs output and error logs of the \acp{rp}.

{\bf Limitation of RP output for studying impact.}
Solely studying the plain output is limited; during validation failures, the \acp{rp} exclude failed objects from the output and subsequently erase all information on these objects, as well as the scope and impact of the failure. 
For example, consider an RPKI repository with an expired TLS certificate.
The \ac{rp} will fail validation of that certificate and log the error.
However, it will not download data from this repository, making it impossible to quantify the exact impact of the expired certificate on the set of VRPs.

To overcome this limitation, we utilize our crawler with validations disabled, running it simultaneously with the \acp{rp} to capture a full view of the global RPKI objects. 
Our measurement was conducted on January 5, 2025.

\subsection{Inconsistencies in VRPs}
In most runs, the \acp{rp} yield differing VRPs outputs.
This observation is consistent with prior work~\cite{mirdita2023cure}, which, however, did not analyze the source of these inconsistencies. 
Our study shows that inconsistencies are caused by: (1) frequent changes to RPKI objects and (2) discrepancies between \ac{rp} implementations. Since the crawler provides us with a view of all \ac{rpki} data before validation, we can map which differences are caused by validation inconsistencies and which are simply due to new objects being added or removed between requests by different RPs. 
Our findings show that most inconsistencies are short-lived and caused by natural fluctuations in object counts. 
We do find two causes for persistent validation inconsistency between RPs impacting VPRs: 

{\em First}, we find that enforcement of manifest number increments discards up to five VRPs. 
This check is only enforced by Routinator and rpki-client, not Fort, adding these VRPs only to the Fort VRPs. 
We observe that the number of VRPs impacted by this enforcement fluctuates. 
Investigating the objects that create this inconsistency, we find that they do not always violate the requirement; the manifest number is incremented only in some updates, not in all.
Since this is an incremental validation, i.e., the check is only conducted when using the incremental update algorithm over deltas, it leads to inconsistencies between snapshot updates and delta updates. 
This makes the detection of this inconsistency even more difficult for an operator, as they might issue the new object, update over snapshot and get the correct VRPs output, whereas an update over delta would fail and exclude the VRPs. 
This breaks with the design philosophy of \ac{rrdp} that applying snapshot or delta updates should always lead to the same deterministic state.

The {\em second} persistent inconsistency occurs in Routinator, resulting in 11 missing VRPs entries. 
Routinator is the only RP that discards URIs referred to as \textit{Dubious hosts}: URIs of unexpected structure, e.g., those including a port number. 
We find that three repositories are not downloaded by Routinator because of the port numbers in their URIs.

\textbf{The issue with inconsistencies.}
Our measurements show that most inconsistencies we identified in RFCs and implementations are currently not present in real-world objects; even the Amazon prefix case was fixed after disclosure and does not currently lead to real-world inconsistencies. 
Our findings are still important to address to ensure resilience of the architecture:
Any of the identified issues could be triggered if an operator makes benign changes to their objects, e.g., adding an additional extension or changing how they use \textit{withdraw} in RRDP, in compliance with the RFC. 
Depending on the test setup used by the operator, the resulting inconsistency can go unnoticed, as with Amazon, leading to a silent downgrade of security for protected resources. 
To ensure that such issues are mitigated \textit{before} they reach production, it is vital to improve RFC quality and address the flaws identified in this paper. 

\section{Factors Introducing the Issues into RPKI}\label{sc:discussion}

A number of principles and decisions have shaped RPKI design in its present form. 
We discuss how the design decisions have led to systematic issues that we find in this work.
We also explain \textit{why} RPKI remains functional and effective today, and how future developments could compromise this.

{\bf Tradeoff of building on existing technologies.} The design of RPKI builds on top of the existing concepts of PKI and applies them to a new use case.
Accordingly, RPKI heavily re-uses existing technologies.
Resource delegation and validation are realized using X.509 certificate validation chains and \acp{crl}.
HTTPS and rsync serve as transport channels for exchanging files.
Existing object formats like \ac{cms}, XML, ASN.1, and DER are used to structurally represent data. 
The application of these existing solutions to RPKI has trade-offs. 
The \ac{rpki} specification requires heavy adaptation and patching of the utilized technologies to fit the \ac{rpki} use-case.
For example, the X.509 certificate structure was designed to be expressive to ensure that certificates are self-sustaining and that distribution does not rely on services like \acp{pp}.
RPKI does not use self-sustaining certificates, as it has a rigid repository structure without out-of-band distribution.
Moreover, X.509 is designed to be very flexible and extensible.
Since RPKI only has a fixed use-case and distribution (hierarchical \acp{pp}), flexibility and extendability are not required (\rfc{6487}).
The RPKI specification acknowledges this and mandates strictness; allowing only specific object formats and disallowing optional and unknown fields (\rfc{6487}), which leads to significant complexity for implementations enforcing this.

Similar observations hold for other adopted technologies.
RPKI uses HTTPS without requiring certificate validation (\S\ref{sc:stricthttp}); and ASN.1 encoding rules are limited to DER (\S\ref{sc:derenc}).
These adaptations and ``tweaks'' introduce complexity to RPKI implementations.
Instead of facilitating the re-use of existing implementations, the substantial changes and new requirements limit the re-usability of existing libraries, as acknowledged by \rfc{6487}, and require \ac{rpki} software to re-implement large parts.
As evident in \S\ref{sc:rfc-analysis}, this is challenging and leads to inconsistencies and vulnerabilities. 
Developers must not only implement and maintain the overhead complexity but also navigate complex interactions of the utilized technologies within \ac{rpki}. 
Examples of how this can lead to issues include implementation vulnerabilities in interacting with HTTPS URIs, rsync URIs, DNS, and X.509 processing~\cite{van2022rpkiller,hlavacek2022behind,mirdita2023cure}. 

{\bf Single point of failure in a distributed system.} 

\ac{rpki} requires RPs to retain an up-to-date view of current data, necessitating a high degree of availability of PPs. 
\rfc{6480}[8] says that \ac{rpki} as a \emph{distributed} repository ``should be inherently resistant to denial-of-service attacks.''
In practice, the way availability is handled in \ac{rpki} renders it vulnerable.
Once a \ac{pp} is unavailable, \acp{rp} can no longer determine whether its \ac{crl} has changed and must not retrieve nor validate its entire sub-tree (\rfc{9286}[6]); even benign misconfigurations or just a single file missing are specified as a failure condition.
Effectively, every unavailable repository in the validation path is a single point of failure for its entire sub-tree.
Compared to other distributed repositories like DNS, \ac{rpki} does not reduce system load by distributing its data.
Every \ac{rp} must walk the full RPKI tree for each validation run.
This makes \ac{rpki} unique among decentralized security infrastructures, as its operations depend on both trust in \emph{and} availability of the anchors.

{\bf Limited impact due to low RPKI adoption.} Our study finds that many problematic design choices of RPKI are currently masked by operational realities. 
Compared to $\sim$45K CAs currently participating in RPKI, the number of self-hosted repositories remains small, with only 100 active instances. 
This low number masks the limited scalability caused by deep and slow repository chains.
Moreover, few active use-cases apart from \acp{roa} are widely deployed yet in RPKI.
ASPAs, BGPsec, \acp{gbr}, and \rfc{8360} validation see little to no use.
Thus, we see low object diversity and little divergence from specifications.
Most of the issues we illustrate in §\ref{sc:rfc-analysis} do not yet occur in real-world objects.
Fixing these issues before they manifest in deployments is, however, essential to ensure RPKI can scale to a resilient and secure architecture with diverse use-cases. 

Existing operational trends indicate a change in the infrastructure that will gradually unmask specification issues.
We measure that the number of repositories has been steadily growing, reaching 100 repositories as of 2025, up from 93 one year ago, which is consistent with observations from previous work~\cite{su2024drr}.
While current divergences are mostly benign, they indicate an increase with more repositories.
Continuing adoption will lead to growth in interoperability issues and implementations defying specifications to account for real-world diversity, e.g., as discussed in \S\ref{sc:derenc}. 

\section{Mitigations and Recommendations}\label{sc:mitigations}

{\bf Test cases for inconsistencies in software.}
Investigating RPKI implementations, we find that unit tests are used to ensure the correct functionality of the software.\footnote{\url{https://github.com/NICMx/FORT-validator/blob/main/test/rrdp_test.c}}

To facilitate additional inconsistency testing, we open-source our implementation-specific tests.

{\bf Minimizing standard complexity.}
\ac{rpki} specifications explicitly state that they prioritize the minimalism of objects over extensibility (\rfc{6487}) and are only intended for one very specific use-case (\rfc{9255}).
Yet, as we show in this work, the design of \ac{rpki} has significant complexity.
We recommend that the design of new RPKI standards should strongly assess the necessity of enforced checks for the respective use-case.
Validation should only be strictly enforced when \textit{necessary} for security or operation and applied leniently in other cases.
To prevent the divergence of objects, the standard should be stricter on object issuers than on RPs. 

{\bf Inconsistency notification.} 
We develop a web service that continuously searches for the issues explored in this work and aggregates information on erroneous or inconsistently handled objects or unreachable repositories (\url{https://rpki-notify.site}).
Stakeholders may register to be notified when new problems occur. 
Additionally, we utilize the contact information in existing \acp{gbr}.
According to \rfc{8897}[4.4], \acp{rp} are expected ``to use the information from \acp{gbr} to contact the maintainer of the \ac{pp} where any stale/expired objects were encountered.''
As of today, this is not implemented in any of the \acp{rp}.
Our service closes this implementation gap and, for the first time, makes use of the \acp{gbr} for their intended purpose by informing affected parties about the problematic objects we detect.
We hope that by offering this service, we further encourage the deployment and usage of \acp{gbr} in \acp{pp}.

{\bf Paving the road for LLMs.}
LLMs offer automated techniques to reduce manual effort in identifying issues.
However, recent work on LLMs found that they remain unreliable for reasoning on complex tasks, like RFC issues~\cite{li2025generation}. 
We thus do not employ LLMs to find and evaluate issues \textit{by themselves}.
Instead, based on the results presented in this work, we explore how LLMs could assist human researchers in detecting and avoiding future issues.
We build an easily customizable LLM-assisted pipeline for identifying candidate sections of specification problems observed in our work and extracting important context for these sections, allowing researchers and RFC authors to better assess potential problems.\\
\indent (1) 

Multiple observed problems (e.g., 1,\,2,\,10) stem from conflicts between sections that reference or update each other.
To identify potential conflicts, we implement an algorithmic routine to contextualize every RFC section by finding all incoming and outgoing references using regular expressions and collecting the referenced sections.
Additionally, we parse all RFC headers and construct the graph of updates and obsoletions.
We then use section titles and references to specific sections to identify which old sections are explicitly updated or replaced by the new RFC.
This automation yields a set of specific RFC sections relevant to analyzing a reference between RFCs.

(2) 
Problems 5, 6 and 9 arise from lacking or conflicting definitions for keywords, e.g., \textit{ASId}. 
To assist in identifying future issues like this, we utilize LLMs.
We use \textit{GPT-4o} by OpenAI for all our evaluations; however, our pipeline supports different models and APIs.
First, we algorithmically identify all sections from the RFCs containing the keyword in question through a regular expression. 
Then, we pass all sections one-by-one to the LLM and task it with deciding whether the section contains a textual or formal (e.g., ASN.1) definition of the term and, if so, to return the definition in the section.
Using this approach, we were able to find all definitions of the term \textit{ASId}, as well as its type and contents, across the various documents, enabling the detection of the mismatch between the expected contents and the overly broad ASN.1 type {INTEGER}.

(3)
Problems 7 and 8 result from significant complexity in the RFCs.
Interdependent requirements distributed over multiple sections/RFCs, lacking proper references, lead to implementers overlooking conflicts and unspecified (corner) cases. 
We implement an LLM-based search functionality to extract the inter-dependencies.
When given a description of a specific requirement to extract, like `limits on the number of validated children repositories,` the tool iterates over all RPKI RFC sections and extracts relevant texts. 
The text can then be manually checked to identify problems. 
In our evaluation, this approach successfully identifies the recommendation to use a maximum chain length from \rfc{6491}.

The code for our tools is available on GitHub.
Further examples of how our tooling could aid a similar RFC analysis can be found in Appendix~\ref{app:automated-analysis}.

\section{Conclusion}\label{sc:conclusion}

Developing implementations from standards requires the interpretation of specification text into code and configurations. Defining requirements is a tradeoff: On the one hand, standards should provide exact details for implementations. On the other hand, they should still leave room for flexibility to allow for competitive implementations. 
Flaws or vagueness in standard requirements not only create hurdles for implementers developing the software but also expose adopters of the standard-compliant technology to attacks.\\ 
\indent This work provides the first systematic analysis of vulnerabilities in RPKI standards and their propagation into implementations and operational deployments. 
Our findings demonstrate that many security and stability issues in RPKI stem not from implementation mistakes but from flaws in the RFC specifications, such as vagueness, conflicting requirements, and underspecified behavior. 
By combining impact-driven specification analysis, differential fuzzing, and real-world validation, we reveal deep-rooted issues and their concrete impact on Internet routing security.\\
\indent In addition to reducing RPKI complexity and improving the resilience of RPKI deployments on the Internet, our research also offers a transferable approach for evaluating other complex, multi-RFC protocols. 
Future work can build on our open-source tooling and datasets to extend this analysis to other protocols and formalize specification testing workflows to support the development of robust standards.\\
\indent To address identified flaws, we are actively working with SIDROPS members to develop fixes to flawed RFC sections. 

\section{Ethics Considerations}\label{sc:ethics}

We ensure that our research is ethical by following best practices for research in network and software security~\cite{bishop2008ethical,dittrich2012menlo,sweeney2015sharing}.
All testing, evaluations and fuzzing activities were performed in controlled environments, isolated from live production systems.
We limit crawler requests to 1/min, as outlined by RRDP specification RFC 8182.
We disclosed all discovered vulnerabilities and inconsistencies to the developers of RPKI software, and two CVEs have been assigned.

\section{LLM Usage Considerations}

The methodology used for this paper does not include any LLM usage.
None of the results presented in our analysis (\S\ref{sc:rfc-analysis}) have been evaluated by LLMs; all identified flaws were manually checked by us, mapped to the RFCs, built into a test-case, and manually run against the implementations to understand how all reacted.
Datasets included as artifacts have been compiled without the use of LLMs.

Building on the results of the methodology and analysis, we develop an LLM tool (referred to in the introduction and \S\ref{sc:mitigations}) to assist future work in analyzing RFCs based on the results obtained from our study.
In our exploration, we limited the number of queries to those necessary to gain insights into the capabilities of LLMs in the tasks explored.

\section*{Acknowledgments}
This research work was supported by the National Research Center for Applied Cybersecurity ATHENE. ATHENE is funded jointly by the German Federal Ministry of Research, Technology and Space and the Hessian Ministry of Science and Research, Arts and Culture.

\bibliographystyle{IEEEtran}
\bibliography{NetSec}

\appendices

\section{Fuzzer Design}\label{app:fuzzer}
To test RPKI implementations we need to simulate a real world RPKI setup. For this, the fuzzer creates \ac{rpki} repositories and assigns each a \ac{roa} with a unique \ac{as} number and IP prefix for fingerprinting. The fuzzer ensures objects are validly structured and encoded to allow objects to pass through parsing and test validation code. To improve deeper penetration of the RP code, the fuzzer guides mutations with coverage guidance. After creating test inputs, the fuzzer executes the \acp{rp} and analyzes their output. 
We implement a parsing module that reads the output logs of the \acp{rp}, and maps error messages to the objects that caused the inconsistency.  
We build a new de-duplication module that aggregates inconsistencies that can be traced back to the same cause. This is necessary, as after one minute of running the fuzzer, already over 99\% of reported inconsistencies are duplicates, requiring substantial manual analysis to filter for new unique findings. Implementing de-duplication is not trivial, as RP log messages follow non-deterministic formats and different variations of error messages can map to the same underlying issue. The module uses the input object type, the pattern of \acp{rp} that accept/reject the object, the processed error messages, and the code lines that results in the error condition to de-duplicate.
We run the fuzzer 3 times for 1h on all RPKI object types and log all new unique inconsistencies. 
We determine the runtime to be sufficient, as we observe no new inconsistencies in longer runs. 
Fuzzing is done on the following versions: Routinator 0.14.2, Fort 1.6.6., rpki-client 9.5.

\section{Automated Analysis}
\label{app:automated-analysis}

We conduct a number of experiments to illustrate how the application of our LLM pipeline could facilitate further research in terms of contextualizing and tracking requirements across RFCs.

\textbf{\findingseravail{}.}

During our analysis, we found that the requirements regarding the availability of RPKI services are spread across a significant number of RFCs, rendering it difficult to obtain a full overview.
We task the LLM with extracting concrete claims regarding the availability of the RPKI, of which we find 10 text passages.
The LLM correctly includes the sections in the \ac{cp} \rfc{6484}, provisioning \rfc{6492}, and \ac{dos} guidelines in \rfc{6487}[6] that we analyzed in Section~\ref{sc:seravail}.
This demonstrates how the LLM pipeline can be used to help keep track of requirements spread out across RFCs.

\textbf{\findingderenc{}.} 
In Section~\ref{sc:derenc}, we find that the RFCs mandate DER encoding and do not require support for CER encoding.
However, ``CER'' is mentioned multiple times within the RPKI RFCs. 
To test our LLM pipeline, we search for all sections containing the keyword ``CER'' and task the LLM with identifying any support requirements.
We find 13 occurrences, none of which the LLM wrongly classifies as support requirements in the \ac{rpki}.

\textbf{\findingextensions{}.}
In \rfc{5280} and \rfc{6487}, we observe the keywords ``differs'' and ``relates'' being used to describe differing requirements, leading to the complexity issue described in Section~\ref{sc:extensions}.
We explore how our LLM tool can be used to investigate potentially recurring patterns.
We perform a keyword search for ``differs'' and ``relates,'' yielding 7 and 3 occurrences, respectively.
We let the LLM decide whether the keywords are used to describe varying or conflicting requirements, and it correctly concludes that none of the occurrences match the observed pattern.

\section{RPKI RFCs}\label{app:rpki-rfcs}

This section provides a full description of the \ac{rpki} RFCs and their categorization.

\textbf{Informational} RFCs give general recommendations and provide extensive descriptions for stakeholders.
Their contents are generally non-binding to implementers.
\rfc{6480} provides an introduction of the structure, operations and purpose of the \ac{rpki}.
\rfc{6483} describes the semantics of \acp{roa} and how route validity is determined on the basis of \ac{roa} payloads.
\rfc{6907, 8897} contain implementation and deployment advice.
\rfc{7128, 8488} discuss specific implementations.
\rfc{8211} describes threats to the \ac{rpki} from adversarial \acp{ca}. 

\textbf{\ac{rpki} Operations / PKI Structure.}
Central to the \ac{rpki} operations is the \rfc{6484} Certificate Policy which specifies how certificates must be issued, distributed, and used.
\rfc{6489, 6916} describe the transition procedures between key pairs and algorithms, respectively.
\rfc{6491} lists \ac{rpki} objects issued by IANA.
\rfc{8416} \ac{slurm} is a mechanism for a \ac{rp} to specify local exceptions to \ac{rpki}.
\rfc{9319, 9455} are updated guidelines on issuing \acp{roa}.

\textbf{Repository Object Format and Validation.}
RFCs in this category define specific objects in the \ac{rpki}, their syntax, semantics, use, and validation.
The repository structure and distribution of objects in \acp{pp} is defined in \rfc{6481}.
The profiles of certificates and \acp{crl} in the \ac{rpki} are defined in \rfc{6487}, and \rfc{9829} and build upon the general X.509 certificate profile in \rfc{5280}.
They make use of the X.509 IP address and AS identifier delegation extensions of \rfc{3779}.
\Ac{rpki} signed objects, such as \acp{roa} (\rfc{6482}, since obsoleted by \rfc{9582}), manifests (\rfc{6486}, since obsoleted by \rfc{9286}), and \acp{gbr} (\rfc{6493}) are the content of a \ac{cms} (\rfc{5652}) signed object, whose template is defined in \rfc{6488} and \rfc{9589}.
The \ac{tal} serves the purpose of linking to the trust anchor CA certificate and is defined in \rfc{8630}, obsoleting the former versions \rfc{7730} and \rfc{6490}.
\rfc{6485}, replaced by \rfc{7935}, defines the algorithms and key sizes in use in the \ac{rpki}.
\rfc{7318} amends \rfc{6487} by permitting \ac{cp} Qualifiers, \rfc{9829} clarifies CRL number handling in \rfc{6487}.
\rfc{8360} ``Validation Reconsidered'' introduces an alternative validation algorithm and new identifiers for use within certificates.

\textbf{BGPsec} (\rfc{8205, 8206}) enables path validation of BGP UPDATE messages and makes use of the \ac{rpki}.
RFCs that make changes to the \ac{rpki} are \rfc{8208}, obsoleted by \rfc{8608}, which amends the algorithms used for BGPsec in \ac{rpki}, and \rfc{8209}, which defines the BGPsec router certificate among others.

\textbf{Production Services.}
RFCs in this category include protocols that do not involve \acp{rp}, instead cover \ac{pp} services.
A provisioning protocol for the \ac{rpki} is defined in \rfc{6492}, defining the interaction between \ac{ca} and subject.
\rfc{8183} describes how a publication relationship can be established out-of-band between a \ac{ca} and a repository operator; \rfc{8181} specifies the publication protocol.

\textbf{Transport Protocols.}
\acp{rp} make use of rsync and the \ac{rrdp} (\rfc{8182, 9674, 9697}) to retrieve objects from the \ac{rpki} distributed repository.
\rfc{6810, 8210} define the \ac{rtr}, \rfc{6945} the monitoring of \ac{rtr}.

\textbf{Validation in Routers.}
The validation of BGP prefix origin announcements in routers using \ac{roa} information is specified in \rfc{6811} and \rfc{7115}.
\rfc{8481} adds clarifications, \rfc{9324} provides fixes to excessive route refreshing, and \rfc{8893} defines \ac{rov} for routes sent to BGP neighbors. Not included, but closely related to the core set of RFCs, are RFCs of the X.509 PKIX (namely \rfc{3779, 5280, 6818}) and vCard format (\rfc{6530}), which provide the specifications for the RPKI X.509 objects and \acp{gbr}, respectively.
\rfc{7909} provides an application of RPKI for securing \acf{rpsl} objects. 

\rfc{9255} touches on RPKI misuse and forbids the usage of the \ac{rpki} to authenticate non-\ac{rpki}, real-world data.

\section{Repository Crawler}\label{app:crawler}

In contrast to RP software packages, which are publicly available, the software of most \ac{rpki} repositories, like software used by the five \acp{rir}, are closed-source and not publicly available.
We develop the repository crawler to evaluate RFC compliance of \ac{rpki} repository instances.
This serves two purposes:

(1) Evaluate repository instance compliance with non-object related \rfc{8182} requirements that impact \acp{rp}.
We validate repositories based on their availability, \ac{rrdp} message syntax, and HTTPS certificates, since these are the mandatory \rfc{8182} requirements that impact whether the responses can be validated by \acp{rp}.

(2) Collect a complete view of all \ac{rpki} objects published in the distributed repository that serves as a benchmark for the validation in \acp{rp}.
We collect a \emph{ground truth} of published data against which we can compare the validated output of \acp{rp} to estimate how much data is ``lost'' due to issues preventing proper validation.
We achieve this by designing the crawler to accept any repository data, regardless of whether it can be validated.

By running the crawler, we obtain a dataset of properties that the repositories uphold and can correlate these data points to inconsistencies between the \acp{vrp} after \ac{rp} validation.

\textbf{Crawler operation.}
The crawler begins with the \acp{tal} of the five \acp{rir} and recursively downloads all repositories by parsing CA certificates, linking manifests, and retrieving all listed objects.
RPKI crawler supports downloads over both \ac{rrdp} and rsync, preferring \ac{rrdp} as per \rfc{8182}.
It validates, but not enforces transport security requirements as mandated by \rfc{8182}, accepting repositories hosted on non-default ports or with invalid or expired TLS certificates.
For \ac{rrdp} messages, the crawler checks syntax requirements, but forgoes enforcement beyond requiring that objects are generally parseable.
Our measurement is conducted on January 5, 2025.
In the following, we present the evaluation results.

\textbf{Repository availability.}
We crawl 99 \ac{rrdp} notification URLs in the full RPKI tree.
Of these, 10 are permanently offline, meaning the URLs to the repository are included in some certificates Subject Information Access extension, but the address did not respond to HTTPS requests from our crawler for any measurement.
3 repository servers are reachable but serve invalid notifications, like nginx error pages, indicating that an HTTPS server is available, but misconfigured.
This leaves 85 repositories that serve RFC-conforming notifications, listing 85 snapshots and 4581 delta files.

\textbf{\ac{rrdp} file syntax.}
In \rfc{8182}[3.5], there are explicit MUST-requirements on the syntax of notification, delta, and snapshot files.
We evaluate all repository files, checking the XML namespace, US-ASCII encoding, version, amount of snapshot entries, and contiguous deltas sequence.

The evaluations show a consistent formatting of \ac{rrdp} files in repository servers.
We parse 86 notification which include 86 snapshot files, and 4573 delta files. All XML files use the correct encoding, version, have exactly one snapshot, and obey the sequencing rules for delta entries.
However, no single file has the correct XML namespace parameter in the root element, indicating that implementers have deliberately deviated from specification, omitting the namespace to reduce the size of RRDP files.
As the namespace is not used for disambiguating XML files, the omitted value does not impact RP operation, however, only because \acp{rp} forgo enforcing the namespace in violation of \rfc{8182}.

\textbf{\ac{rrdp} HTTPS certificates.}
Data published in the RPKI repository is fully signed.
By design, repository servers and communication channels to \acp{rp} are untrusted.
Nevertheless, \ac{rrdp} utilizes HTTP over TLS (HTTPS) for downloading RPKI data.
\rfc{8182}[4.3] has the explicit requirement that \acp{rp} should validate TLS certificates obtained from the servers, but they must not reject data based on this.
Furthermore, several guidelines on TLS certificates are listed, against which we evaluate the TLS certificates of repository servers: DNS-ID identifiers should be present, DNS names should not contain the wildcard ``*,'' and a Common Name field may be present but should not be used for authentication.
We run the crawler to collect all HTTPS certificates of available RPKI repository servers.
Of the 65 obtained certificates, 62 are valid.
Two certificates had been issued by LetsEncrypt, but have since expired, and one certificate is self-signed and lacks DNS-IDs.
Moreover, 16 ($25\%$) have DNS wildcard characters, and one has no common name.

\section{Vulnerabilities}\label{app:vuln}
During our work we identified two vulnerabilities and received two CVEs for our findings (CVE-2025-0638, CVE-2024-56375).

\textbf{Routinator.} The Routinator vulnerable enables DoS of the implementation by uploading a malformed RPKI object to a repository. The vulnerability results from a UTF-8 parsing error that led to an unhandled exception in the code. Specifically, if a manifest entry contains the UTF-8 continuation character without a proper termination in the next character, Routinator will crash.

\textbf{Fort.} The Fort vulnerability results from a missing bounds check in the manifest entry iteration not considering the case of an empty list. If an empty list was provided, the loop check overflowed, and the loop executed with a length of zero, leading to a field accessed in an uninitialized array and crashing the binary.

\section{Meta-Review}

The following meta-review was prepared by the program committee for the 2026
IEEE Symposium on Security and Privacy (S\&P) as part of the review process as
detailed in the call for papers.

\subsection{Summary}
This paper analyzes the Resource Public Key Infrastructure (RPKI) ecosystem to identify how vagueness and inconsistencies in RFC specifications lead to implementation bugs and security vulnerabilities. 

\subsection{Scientific Contributions}
\begin{itemize}
\item Identifies an Impactful Vulnerability
\item Provides a Valuable Step Forward in an Established Field
\end{itemize}

\subsection{Reasons for Acceptance}
\begin{enumerate}
\item \textbf{Identifies an Impactful Vulnerability:} The scope of the study is comprehensive: it spans many RFCs and provides a holistic view of the RPKI landscape.
\item \textbf{Provides a Valuable Step Forward in an Established Field:} The ``impact-driven'' reverse-direction analysis provides a powerful template for analyzing other complex protocol; it goes from implementation errors to specification flaws.
\end{enumerate}

\subsection{Noteworthy Concern} 
\textbf{Causality and Root Cause Attribution:} A significant discussion point was whether implementation divergences are truly caused by RFC flaws or if they represent flexible developer choices that do not necessarily lead to security impacts. The paper must carefully distinguish between these two.

\end{document}